\documentclass[10pt,showpacs,floatfix,letterpaper,amsmath,amsfonts,amssymb,aps,pra,reprint,longbibliography,superscriptaddress]{revtex4-1}

\usepackage{color}
\usepackage{tensor}
\usepackage{graphicx}
\usepackage{amsmath}
\usepackage{epstopdf}
\usepackage{enumerate}
\usepackage[latin1]{inputenc}
\usepackage{bm}
\usepackage{xfrac}
\usepackage[caption=false]{subfig}

\usepackage[pdfpagemode=UseNone,pdfstartview=FitH,colorlinks=true,linkcolor=blue,citecolor=blue,urlcolor=blue]{hyperref}
\usepackage[all]{hypcap}

\newcommand{\ket}[1]{\lvert #1\rangle}           
\newcommand{\bra}[1]{\langle #1\rvert}           
\newcommand{\braket}[2]{\langle #1 \vert #2 \rangle}                                         

\newcommand{\be} {\begin{equation}}			
\newcommand{\ee} {\end{equation}}				

\begin{document}
\title{Measuring a transmon qubit in circuit QED: dressed squeezed states}
\author{Mostafa Khezri}
\email[email: ]{mostafa.khezri@email.ucr.edu}
\affiliation{Department of Electrical and Computer Engineering, University of California, Riverside, California 92521, USA.}
\affiliation{Department of Physics, University of California, Riverside, California 92521, USA.}
\author{Eric Mlinar}
\affiliation{Department of Electrical and Computer Engineering, University of California, Riverside, California 92521, USA.}
\author{Justin Dressel}
\affiliation{Institute for Quantum Studies, Chapman University, Orange, California 92866, USA.}
\affiliation{Schmid College of Science and Technology, Chapman University, Orange, California 92866, USA.}
\author{Alexander N. Korotkov}
\affiliation{Department of Electrical and Computer Engineering, University of California, Riverside, California 92521, USA.}
\date{\today}


\begin{abstract}
Using circuit QED, we consider the measurement of a superconducting transmon qubit via a coupled microwave resonator. For ideally dispersive coupling, ringing up the resonator produces coherent states with frequencies matched to transmon energy states. Realistic coupling is not ideally dispersive, however, so transmon-resonator energy levels hybridize into joint eigenstate ladders of the Jaynes-Cummings type. Previous work has shown that ringing up the resonator approximately respects this ladder structure to produce a coherent state in the eigenbasis (a dressed coherent state). We numerically investigate the validity of this coherent state approximation to find two primary deviations. First, resonator ring-up leaks small stray populations into eigenstate ladders corresponding to different transmon states. Second, within an eigenstate ladder the transmon nonlinearity shears the coherent state as it evolves. We then show that the next natural approximation for this sheared state in the eigenbasis is a dressed squeezed state, and derive simple evolution equations for such states using a hybrid phase-Fock-space description.
\end{abstract}
\maketitle

\section{Introduction}\label{sec:intro}

Qubit technology using superconducting circuit quantum electrodynamics (QED) \cite{Blais2004,Wallraff2004} has rapidly developed over the past decade to become a leading contender for realizing a scalable quantum computer. Most recent qubit designs favor variations of the transmon \cite{Koch2007,Schreier2008,3DTransmon2011,Xmon2013,PreGatemon2015,Gatemon2015,Murch2013} due to its charge-noise insensitivity, which permits long coherence times while also enabling high-fidelity quantum gates \cite{Chow2012,Barends2014,Chow2014} and high-fidelity dispersive qubit readout \cite{Vijay2011,Riste2012,Jeffrey2014} via coupled microwave resonators. Transmon-based circuit operation fidelities are now near the threshold for quantum error correction protocols, some versions of which have been realized \cite{Reed2012,Kelly2015,Corcoles2015,Riste2015}.

The quantized energy states of a transmon are measured in circuit QED by coupling them to a detuned microwave resonator. For low numbers of photons populating the readout resonator, the coupling is well-studied \cite{Blais2004,Koch2007,Gambetta2006} and approximates an idealized dispersive quantum non-demolition (QND) measurement \cite{Braginsky1995-book}. Each transmon energy level dispersively shifts the frequency of the coupled resonator by a distinct amount, allowing the transmon state to be determined by measurement of the microwave field transmitted through or reflected from the resonator. However, nondispersive effects become important when the number of resonator photons becomes  comparable to a characteristic (``critical'') number set by the detuning and coupling strength \cite{Blais2004,Boissonneault2008,Boissonneault2010}; present-day experiments often operate in this nondispersive (or nonlinear dispersive) regime \cite{Jeffrey2014,Slichter2016,Bultink2016,Sank2016}.

In this paper, we analyze and model the nondispersive effects that occur during the ring-up of a readout resonator coupled to a transmon. These effects arise from the hybridization of the resonator and transmon states into joint resonator-transmon eigenstates. While ringing up the resonator from its ground state, the joint state remains largely confined to a single Jaynes-Cummings eigenstate ladder that corresponds to the initial transmon state. As pointed out in Refs.~\cite{Sete2013,Sete2014,Govia2016}, this joint state can be approximated by a coherent state in the eigenbasis (recently named a dressed coherent state \cite{Govia2016}). Here we refine this initial approximation and provide a more accurate model for the hybridized resonator-transmon state.

We numerically simulate the ring-up process for a resonator coupled to a transmon, then use this simulation to develop and verify our analytical model. We find two dominant deviations from a dressed coherent state. First, we show that the ring-up process allows a small population to leak from an initial transmon state into neighboring eigenstate ladders, and find simple expressions that quantify this stray population. Second, we show that the transmon-induced nonlinearity of the resonator distorts the dressed coherent state remaining in the correct eigenstate ladder with a shearing effect as it evolves, and show that this effect closely approximates self-squeezing of the dressed field at higher photon numbers. We then use a hybrid phase-Fock-space method to find equations of motion for the parameters of an effective \emph{dressed squeezed state} that is formed during the ring-up process. Our improved model is satisfyingly simple yet quite accurate.

To simplify our analysis and isolate the hybridization effects of interest, we restrict our attention to a transmon (modeled as a seven-level nonlinear oscillator) coupled to a coherently pumped but non-leaking resonator (using the rotating wave approximation). The simplification of no resonator leakage may seem artificial, but it is still a reasonable approximation during the resonator ring-up and it is also relevant for at least two known protocols. First, the catch-disperse-release protocol \cite{Sete2013} encodes qubit information into resonator states with minimal initial leakage, then rapidly releases the resonator field to a transmission line. Second, a recently proposed readout protocol \cite{Govia2014} similarly encodes qubit information into bright and dark resonator states with minimal leakage, then rapidly distinguishes them destructively using Josephson photomultipliers \cite{Chen2011}. Our dressed squeezed state model should describe the ring-up process of these and similar protocols reasonably well. Additional effects arising from a more realistic treatment of the resonator decay will be considered in future work.

Our assumption of negligible resonator damping automatically eliminates qubit relaxation (and excitation) due to the Purcell effect \cite{Esteve1986,Houck2008,Boissonneault2008,Sete2014}, which in the present-day experiments is often strongly suppressed by Purcell filters \cite{Reed2010,Jeffrey2014,Sete2015}. We also neglect energy relaxation and dephasing of the qubit (thus also eliminating dressed dephasing \cite{Boissonneault2008,Boissonneault2009}).

We note that squeezing of the resonator field may significantly affect fidelity of the qubit measurement \cite{Mallet2009,Krantz2016}, which can be either increased or decreased, depending on the squeezing axis direction. A significant improvement of the fidelity due to self-developing quadrature squeezing was predicted for the catch-disperse-release protocol \cite{Sete2013}. (An extreme regime of the self-developing squeezing, with revival and formation of ``cat'' states was experimentally observed in \cite{Kirchmair2013}.) The use of a squeezed input microwave for the qubit measurement was analyzed in \cite{Barzanjeh2014}. A Heisenberg-limited scaling for the qubit readout was predicted for the two-resonator measurement scheme based on two-mode squeezed microwave in \cite{Didier2015}. The significant current interest in various uses of squeezed microwave states \cite{Barzanjeh2014,Didier2015,Govia2014b,Puri2016} is supported by a natural way of producing them with Josephson parametric amplifiers \cite{Castellanos2008,Bergeal2010,Eichler2011,Flurin2012,Murch2013b,Toyli2016}. All this motivates the  importance of studying squeezed microwave fields in superconducting circuits containing qubits.

The paper is organized as follows. In Sec.~\ref{sec:model} we describe the resonator-transmon system and how the numerical simulations are performed. In Sec.~\ref{sec:DCS} we discuss the dressed coherent state model and focus on analyzing the inaccuracy of this model relative to numerical simulation. We quantify two deviations from the dressed coherent state model: stray population leakage to incorrect eigenstate ladders (Sec.\ \ref{sec:eigen}), and distortion of the remaining dressed state during evolution into a dressed sheared state (Sec.\ \ref{sec:shearing}). In Sec.~\ref{sec:squeezed}, we prove that a dressed sheared state approximates a dressed squeezed state and then derive hybrid phase-Fock-space evolution equations for such states. Comparison with the simulation results shows that the accuracy of the dressed squeezed state approximation is much better than accuracy of the dressed coherent state approximation. We conclude in Sec.~\ref{sec:conclusion}. In the Appendix we show that, somewhat unexpectedly, dressed coherent states and dressed squeezed states are practically unentangled despite the strong entanglement of the dressed Fock states from which they are composed.

\section{Model}\label{sec:model}

\begin{figure}[t]
	\begin{center}
		\includegraphics[width=8cm]{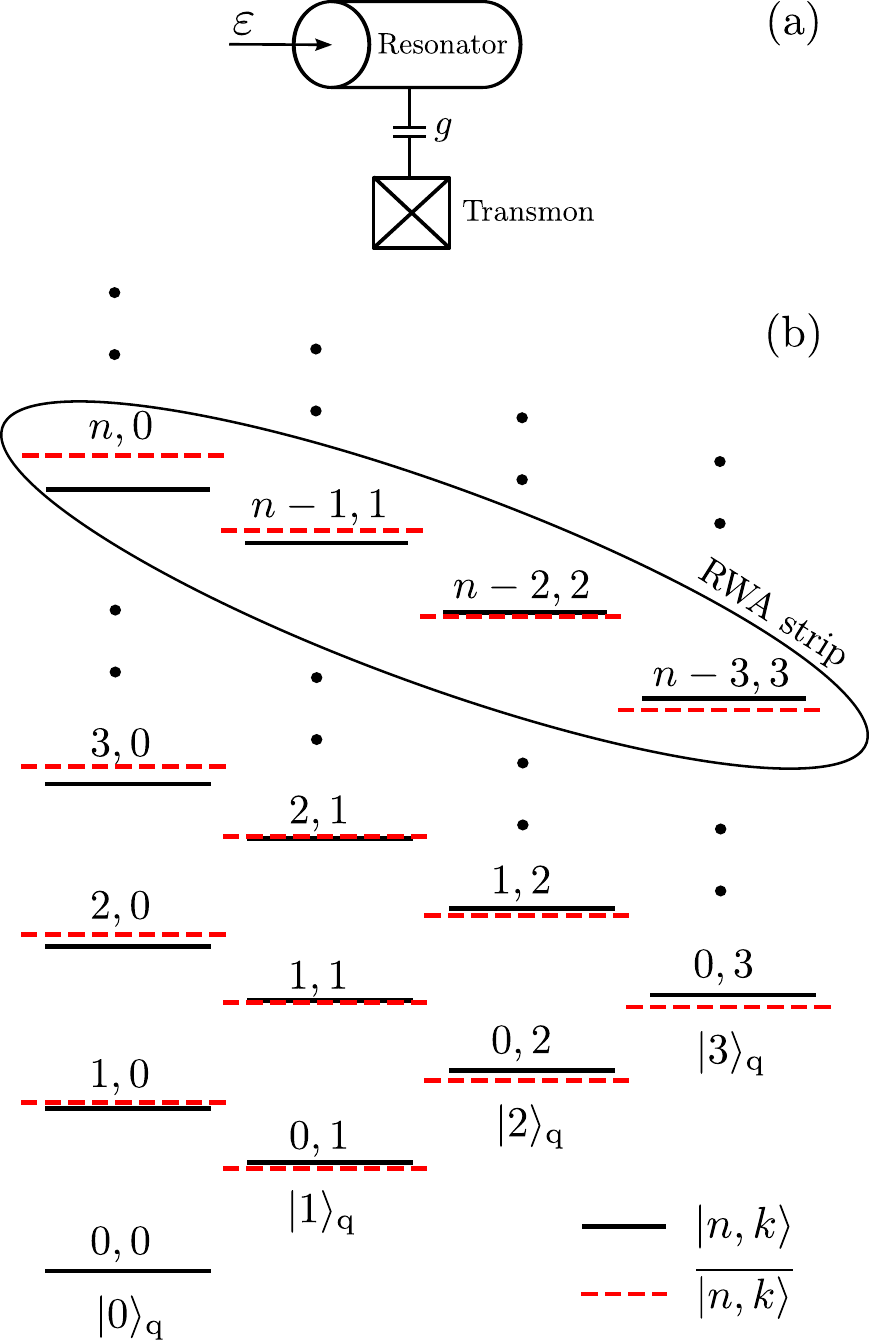}
    \end{center}
    \caption{(a) Considered system: a transmon coupled to a pumped resonator. The resonator damping is neglected, since we focus on the resonator ring-up and/or setups with a tunable coupler. (b) Jaynes-Cummings ladder of states. Bare states are shown by solid black lines. Eigenlevels are shown by red dashed lines. When $n\agt n_c$, the eigenlevels are significantly different from bare levels.} \label{fig:schematic}
\end{figure}

Following the circuit QED paradigm of measurement \cite{Blais2004}, we consider a transmon coupled to a detuned readout resonator (Fig.\ \ref{fig:schematic}). We do not simplify the transmon to a 2-level qubit, but instead include the lowest 7 energy levels confined by the cosine potential of the transmon. Though the transmon eigenstates may be written explicitly as Mathieu functions \cite{Koch2007,Devoret2004}, we have checked that a perturbative treatment of the transmon as an approximate oscillator with quartic anharmonicity \cite{Koch2007} is sufficiently accurate for our purposes. We assume a transmon-resonator coupling of Jaynes-Cummings type \cite{Jaynes1963}, using the rotating wave approximation (RWA) for simplicity. (Notably, this approximation  fails at very high photon numbers, leading to important effects \cite{Sank2016}.)

\subsection{Pumped resonator-transmon Hamiltonian}

In our model the resonator Hamiltonian is
\begin{align}\label{eq:H-resonator}
	H_\mathrm{r} = \omega_\mathrm{r}\,a^\dagger a = \sum_{n,k}n\,\omega_\mathrm{r}\,\ket{n,k}\bra{n,k},
\end{align}
with $\hbar = 1$, bare resonator frequency $\omega_\mathrm{r}$, lowering (raising) operator $a$ ($a^\dagger$) for the resonator mode satisfying $[a,a^\dagger]=1$, and resonator index $n = 0,1,\ldots$ for successive energy levels. For completeness we included the transmon index $k = 0,1,\ldots,6$ for the 7 lowest levels to emphasize the matrix representation in terms of the joint product states $\ket{n,k} \equiv \ket{n}_{\rm r}\otimes\ket{k}_{\rm q}$ for the bare energy states.

Similarly, the transmon Hamiltonian has the form
\begin{align}\label{eq:H-qubit}
	H_\mathrm{q} &= \sum_{n,k}E_k\,\ket{n,k}\bra{n,k}, \\
    \label{eq:H-qubit-freq}
    E_k & = E_0 + \omega_\mathrm{q} k - \eta \,\frac{k(k-1)}{2}.
\end{align}
The dominant effect of the nonlinearity of the cosine potential for the transmon is the quartic anharmonicity $\eta \equiv \omega_{10}-\omega_{21} > 0$ of the upper level frequency spacings relative to the qubit frequency $\omega_\mathrm{q} \equiv \omega_{10}$, where each frequency $\omega_{k\ell} \equiv E_k - E_\ell$ denotes an energy difference. At this level of approximation, the transmon has the structure of a Duffing oscillator with a linearly accumulating anharmonicity $\omega_{(k+1)k} = \omega_\mathrm{q}- k\,\eta$. [This approximation is sometimes extended to an infinite number of levels, $H_\mathrm{q} = E_0 + \omega_\mathrm{q}\, b^\dagger b - (\eta/2)\, b^\dagger b(b^\dagger b -1)$ \cite{Cross2015}, with an effective oscillator lowering (raising) operator $b$ ($b^\dagger$) satisfying $[b,b^\dagger]=1$, but we explicitly keep only the 7 lowest levels here.]

The excitation-preserving interaction (within RWA) is
\begin{equation}\label{eq:H_I}
	H_\mathrm{I} = \sum_{n,k}\, g\,\sqrt{n(k+1)}\,\ket{n-1,k+1}\bra{n,k} + \mathrm{H.c.},
\end{equation}
where $g$ is the coupling strength between levels $|0,1\rangle$ and $|1,0\rangle$. As in Ref.~\cite{Koch2007}, we neglect the effects of the anharmonicity $\eta$ in the coupling for simplicity. [Extending this coupling to an infinite number of transmon levels yields $H_I = g\,(ab^\dagger + a^\dagger b)$.]

Finally, the Hamiltonian for coherently pumping the resonator with a classical field $\varepsilon(t)\, e^{-i\omega_\mathrm{d} t}$ is (within RWA)
\begin{align}\label{eq:H_d}
	H_\mathrm{d} &= \varepsilon(t)\, e^{-i\omega_\mathrm{d} t}\,a^\dagger + \varepsilon^*(t)\, e^{i\omega_\mathrm{d} t}\,a \nonumber \\
    &= \varepsilon(t)\, e^{-i\omega_\mathrm{d} t}\sum_{n,k}\sqrt{n+1}\,\ket{n+1,k}\bra{n,k} + \mathrm{H.c.},
\end{align}
where $\varepsilon(t)$ is a complex envelope for the drive.

Combining Eqs.~\eqref{eq:H-resonator}--\eqref{eq:H_d} into the total Hamiltonian $H = H_\mathrm{r} + H_\mathrm{q} + H_\mathrm{I} + H_\mathrm{d}$, and rewriting it in the rotating frame of the drive frequency $\omega_\mathrm{d}$ yields
\begin{align}\label{eq:H-rf}
	H_\mathrm{rot} = \sum_{n,k}&\big\{\left[n\,(\omega_\mathrm{r}-\omega_\mathrm{d}) + (E_k-k\,\omega_\mathrm{d})\right]\ket{n,k}\bra{n,k}  \nonumber \\
	&+ g\,\sqrt{n(k+1)}\,\ket{n-1,k+1}\bra{n,k} + \mathrm{H.c.} \nonumber \\
	&+ \varepsilon(t)\,\sqrt{n+1}\,\ket{n+1,k}\bra{n,k} + \mathrm{H.c.}\, \big\}.
\end{align}
This simplified Hamiltonian will be sufficient in what follows to observe the dominant non-dispersive effects that affect the resonator ring-up. Note that we use the rotating frame in numerical simulations, but physics related to Jaynes-Cummings ladders of states is easier to understand in the lab frame, so we will often imply the lab frame for clarity in the discussions below.

\subsection{Numerical simulation and diagonalization}

For numerical simulation, the Hamiltonian in Eq.~\eqref{eq:H-rf} is represented by a $7N\times7N$ matrix using the bare energy basis $\ket{n,k}$, where $N=200$--800 is the maximum number of simulated levels for the resonator. We choose experimentally relevant resonator and transmon parameters, which in most simulations are $\omega_\mathrm{r}/2\pi = 6\,\text{GHz}$, $\omega_\mathrm{q}/2\pi = 5\,\text{GHz}$, $\eta/2\pi = 200\,\text{MHz}$, and $g/2\pi = 100\,\text{MHz}$. For the drive, we change the frequency $\omega_\mathrm{d}$ to be resonant with specific eigenstate transition frequencies of interest (detailed later) and use drive amplitudes typically in the range $\varepsilon/2\pi= 10$--60 MHz.

The hybridization of the joint eigenstates [see Fig.~\ref{fig:schematic}(b)] is significant when the number of photons $n$ in the resonator is comparable to or larger than the so-called critical photon number \cite{Blais2004,Boissonneault2008,Boissonneault2010},
\begin{align}
  n_c = \frac{(\omega_\mathrm{r} - \omega_\mathrm{q})^2}{4g^2}.
\end{align}
For the above parameters $n_c = 25$. This defines the scale at which we expect significant deviations from the ideal dispersive model.

We use the following numerical procedure for identifying the joint hybridized eigenstates $\overline{\ket{n,k}}$ of Eq.~\eqref{eq:H-rf} without a drive---we will distinguish dressed (eigen) states (and operators) from bare states by an overline throughout. After setting $\varepsilon = 0$ to eliminate the drive, the matrix representation of Eq.~\eqref{eq:H-rf} is numerically diagonalized to obtain an initially unsorted list of matched eigenenergy/eigenstate pairs $\{\overline{E}_{n,k}, \overline{\ket{n,k}}\}$ for the qubit-resonator system. The one-to-one correspondence between these pairs and the bare energy/state pairs $\{E_{n,k}, \ket{n,k}\}$ may be found by examining the structure of the RWA interaction Hamiltonian in Eq.~\eqref{eq:H_I}: Since excitation number is preserved, there exist closed subspaces $\{\ket{n,k} : (n+k)=n_\Sigma \}$ with constant excitation number $n_\Sigma = 0,1,\ldots$, which we name \emph{RWA strips} \cite{Sank2016}  [see Fig.~\ref{fig:schematic}(b)]. Crucially, since energy levels repel during interaction and avoid crossing, the order of the eigenenergies within a strip is the same as for bare energies. Thus, for each strip with $n_\Sigma$ excitations we first identify the eigenstates  $\overline{\ket{n,k}}$ that lie within the span of that strip; next, we order the eigenenergies $\overline{E}_{n,k}$ to match the bare energies $E_{n,k}$, which uniquely identifies each hybridized eigenenergy/eigenstate pair. We then set the overall sign of each eigenstate such that it does not flip with changing $n$.
After performing this identification, we construct a basis-change matrix
\begin{align}\label{eq:basisU}
  U \equiv \sum_{n,k} \overline{\ket{n,k}}\bra{n,k}
\end{align}
to easily switch between representations numerically.
Note that without proper identification (sorting) of the eigenstates, the numerical analysis at large photon numbers is practically impossible.

The eigenstates $\overline{\ket{n,k}}$ form the Jaynes-Cummings ladders
of effective resonator levels that correspond to a fixed nominal qubit level $k$. For brevity we will call them {\it eigenladders} of dressed resonator Fock states. Each eigenladder behaves like a nonlinear resonator, with an $n$-dependent frequency
\begin{align}\label{eq:eigenfreq}
  \omega_{\mathrm{r}}^{(k)}(n) &=  \overline{E}_{n+1,k} - \overline{E}_{n,k}.
\end{align}
Note that in this formula both sides are numerically calculated in the rotating frame; however, the equation in the lab frame is the same. Conversion to the lab frame involves adding the drive frequency:  $\omega_{\rm d} + \omega_\mathrm{r}^{(k)}(n)$ for the resonator frequency and  $(n+k)\,\omega_{\rm d} + \overline{E}_{n,k}$ for energy.

At large photon numbers, $n\agt n_c$, each $\overline{\ket{n,k}}$ spans a significant fraction of all bare transmon levels. Nevertheless, as we will see, ringing up the resonator from its ground state with an initial transmon level $k$ will primarily excite the states within the eigenladder corresponding to $k$. This behavior closely mimics that of the ideal dispersive case, where a pump excites the bare resonator states $\ket{n}_{\rm r}$ while keeping the transmon state $\ket{k}_{\rm q}$ unperturbed. However, we will also show that there are small but important dynamical differences between our RWA Jaynes-Cummings model and ideal dispersive coupling in the eigenbasis.

\section{Dressed coherent state model}\label{sec:DCS}

We now define an ideal coherent state in the eigenbasis  \cite{Sete2013,Sete2014,Govia2016} (a dressed coherent state) corresponding to a nominal transmon state $k$ as
\begin{equation}\label{eq:DCS}
	\ket{\alpha}_k = e^{-|\alpha|^2/2} \sum_{n} \frac{\alpha^n}{\sqrt{n!}} \,\overline{\ket{n,k}},
\end{equation}
so that the only difference from the standard coherent state of the resonator is that we use eigenstates instead of the bare states.
Perhaps surprisingly given the eigenstate hybridization, such a dressed coherent state is practically unentangled even for $|\alpha|^2\gg  n_c$, in contrast to what one might initially guess \cite{Govia2016}---see Appendix.

A dressed coherent state is not an eigenstate of the bare lowering operator $a$ of the resonator. Instead, it is an eigenstate of the \emph{dressed} lowering operator \cite{Boissonneault2009,Sete2014}
\begin{align}\label{eq:eigena}
  \overline{a} &\equiv UaU^\dagger = \sum_{n,k} \sqrt{n+1}\,\,\overline{|n,k}\rangle \langle \overline{n+1,k|} 
\end{align}
that removes a collective excitation within the same eigenladder. The parameter $\alpha$ is the expectation value of the dressed lowering operator, $\alpha = {}_k\bra{\alpha}\overline{a}\ket{\alpha}_k$, which will be useful in what follows.

Note that for a dressed coherent state $|\alpha\rangle_k$, $|\alpha|^2$ is not exactly equal to the average number $\bar{n}$ of photons in the resonator. (Instead, $|\alpha|^2 = {}_k\bra{\alpha}\overline{a}^\dagger\overline{a}\ket{\alpha}_k$ is the average dressed excitation number within eigenladder $k$.) However, the difference is very small and will be mostly neglected below, so that we will use $\bar{n}=|\alpha|^2$. In the cases when the difference may be important, we will specify the meaning of $\bar{n}$ explicitly.

\subsection{Model inaccuracy contributions}\label{sec:infidelity-def}

During resonator ring-up, we expect the joint qubit-resonator state to approximate such a dressed coherent state, rather than a bare coherent state as is usually assumed with ideal dispersive coupling. As such, we quantify the fidelity of a numerically simulated state $\ket{\psi}$ compared to a dressed coherent state $\ket{\alpha}_k$ as the overlap
\begin{align}
  F = |\braket{\psi}{\alpha}_k|^2,
\end{align}
where the parameter $\alpha$ is chosen to maximize the fidelity. In practice, we find that an initial guess of $\alpha = \bra{\psi}\overline{a}\ket{\psi}$ is very close to the optimal $\alpha$, producing nearly indistinguishable fidelity.

Note that we can expand a numerically calculated state $\ket{\psi} = \sum_{n,\ell} c_{n,\ell}\,\overline{\ket{n,\ell}}$ as
\begin{align}\label{eq:statesplit}
  \ket{\psi} =  \sqrt{1-P_\mathrm{stray}}\,\ket{\psi}_k + \sqrt{P_\mathrm{stray}}\,\ket{\psi}_\perp,
\end{align}
splitting it into a part $\ket{\psi}_k \propto \sum_n c_{n,k}\,\overline{\ket{n,k}}$ within the ``correct'' eigenladder $k$, and a part $\ket{\psi}_\perp \propto \sum_{n,\ell\neq k}c_{n,\ell}\,\overline{\ket{n,\ell}}$ orthogonal to that eigenladder, where $P_\mathrm{stray} = \sum_{n,\ell\neq k}|c_{n,\ell}|^2$ is the stray population that leaked out of the eigenladder $k$, and both $\ket{\psi}_k$ and $\ket{\psi}_\perp$ are normalized. As such, if we define the overlap fidelity within the correct eigenladder $F_\mathrm{c} = |{}_k\braket{\alpha}{\psi}_k|^2$, then we can write the total fidelity as $F = (1-P_\mathrm{stray})\,F_{\rm c}$, and thus decompose the infidelity
\begin{equation}\label{eq:fid}
	1-F = P_\mathrm{stray} + (1-P_\mathrm{stray})(1-F_\mathrm{c})
\end{equation}
into two distinct sources: (i) the stray population $P_\mathrm{stray}$ outside the correct eigenladder, and (ii) the infidelity $1-F_\mathrm{c}$ compared with a coherent state within the correct eigenladder.

\begin{figure}[t]
	\begin{center}
		\includegraphics[width=8cm]{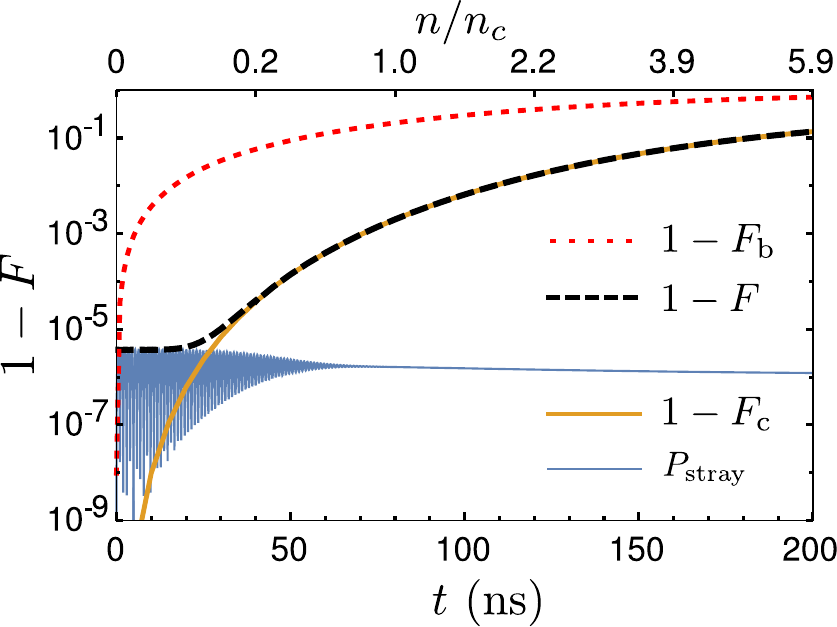}
    \end{center}
    \caption{Infidelity of coherent-state approximations during resonator ring-up. The  infidelity $1-F_\mathrm{b}$ of a bare coherent state (dotted red line) is compared with the infidelity $1-F$ of a dressed coherent state (dashed black line). The latter displays two distinct effects: at short time (and  small photon number $\bar{n}$) the dominant effect is the leakage of a stray population $P_\mathrm{stray}$ (thin solid blue line) out of the correct eigenladder; however, at longer time (and larger $\bar{n}$) the infidelity $1-F_{\rm c}$ of the renormalized state within the correct eigenladder (thick solid orange line)  significantly increases during evolution. Here the system, with parameters $\omega_\mathrm{r}/2\pi=6\,\text{GHz}$, $\omega_\mathrm{q}/2\pi=5\,\text{GHz}$, $\eta/2\pi=200\,\text{MHz}$, $g/2\pi=100\,\text{MHz}$, is resonantly pumped from its ground state $|0,0\rangle$ with a constant drive envelope $\varepsilon/2\pi=10\,\text{MHz}$.} \label{fig:infid}
\end{figure}

To test the infidelity of the dressed coherent state model, we numerically simulate the resonator ring-up with a (sudden) constant drive amplitude $\varepsilon/2\pi = 10$ MHz, and then calculate the infidelity according to Eq.~\eqref{eq:fid} as a function of time, yielding the results presented in Fig.~\ref{fig:infid}. First, we confirm that the infidelity $1-F$ for a dressed coherent state (black dashed line) is typically orders of magnitude better than the infidelity $1-F_\mathrm{b}$ for a bare coherent state (red dotted line); as expected, $1-F_\mathrm{b}$ becomes very significant at $n\agt n_c$. Second, we can clearly separate the effects of the stray population leakage $P_\mathrm{stray}$ (thin solid blue line) from the infidelity $1-F_{\rm c}$ of the renormalized state within the correct eigenladder (thick solid orange line). At short times, the dominant effect is a small (${\sim}10^{-5}$) stray population leakage that rapidly oscillates and then stays approximately constant. (For clarity we do not show oscillations for the black dashed line, showing only the maxima.) However, at longer times the contribution $1-F_{\rm c}$ becomes the dominant source of infidelity (eventually reaching ${\sim}10^{-1}$). In the next two subsections, we quantify these two sources of infidelity in more detail.

\subsection{Infidelity from stray population}\label{sec:eigen}

\begin{figure}[t]
	\begin{center}
 \includegraphics[width=\columnwidth]{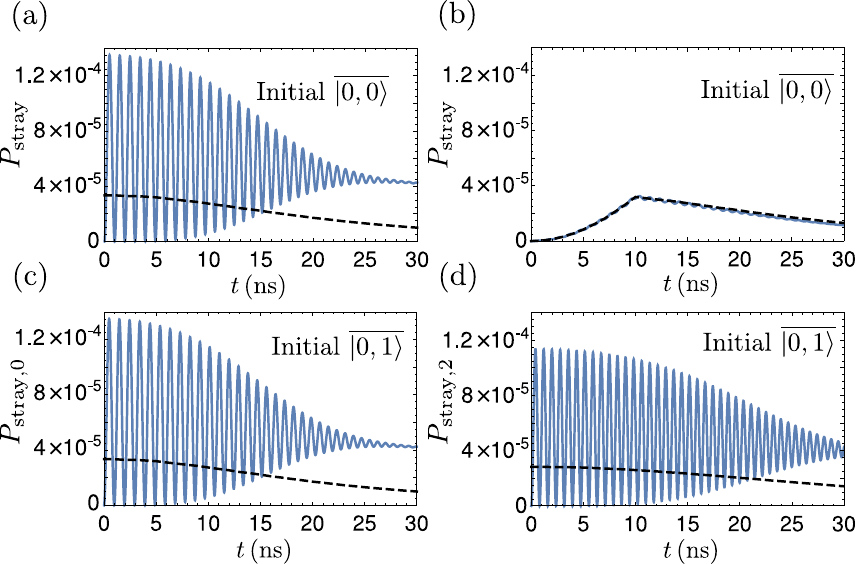}
    \end{center}
    \caption{Solid blue lines: numerically calculated stray population $P_\mathrm{stray}$ as a function of time $t$; dashed black lines: steady-state value $P_{\rm s.s.}(t)$, calculated via Eq.\ (\ref{eq:pstray}). (a) Leaked population in the excited eigenladder $\overline{|n,1\rangle}$ for sudden driving with $\varepsilon/2\pi=60\,\text{MHz}$ from initial ground state $|0,0\rangle$. The oscillations reach an initial maximum of $P_{\rm max}\approx 4 P_{\rm s.s.}(0)$, then dephase to about $P_{\rm s.s.}(t) + P_{\rm s.s.}(0)$, with decreasing $P_{\rm s.s.}(t)$ because of increasing average photon number $\bar{n}$. (b) The same for adiabatic drive $\varepsilon(t)$, linearly increasing for first $10\,\text{ns}$ to the same constant value of $60\,\text{MHz}$. The stray population follows the steady state, which increases for 10 ns because of increasing $\varepsilon (t)$. (c) Sudden driving with $\varepsilon/2\pi=60\,\text{MHz}$  from an initial excited qubit state $\overline{|0,1\rangle}$, showing population $P_\mathrm{stray, 0}$ leaked to the ground-state eigenladder $\overline{|n,0\rangle}$. This case is fully symmetric with (a) since it involves the same pair of transmon levels. (d) The same driving as in (c), but showing leaked population $P_\mathrm{stray, 2}$ of the second-excited eigenladder $\overline{|n,2\rangle}$. The behavior is similar to (c), but involves the next pair of transmon levels. For all panels $\omega_\mathrm{r}/2\pi=6\,\text{GHz}$, $\omega_\mathrm{q}/2\pi=5\,\text{GHz}$, $\eta/2\pi=250\,\text{MHz}$, $g/2\pi=100\,\text{MHz}$, and $\omega_{\rm d}$ is on resonance with the resonator frequency, corresponding to each initial state.} \label{fig:rho-t}
\end{figure}

We now focus on the cause of the stray population outside the correct eigenladder. (Recall that our model neglects qubit energy relaxation, dressed dephasing, and Purcell effect.) Fig.~\ref{fig:rho-t} shows numerical results for different choices of initial state and drive amplitude, produced in a manner similar to Fig.~\ref{fig:infid}, but focusing on shorter times and lower photon numbers, where the stray population is the dominant source of infidelity. Initially, the stray population rapidly oscillates from zero around a steady-state value, then the oscillations damp, after which the stray population continues to slowly decay on a longer time scale. We now provide a phenomenological model that describes this behavior.

A dressed coherent state would naturally be produced by a dressed displacement Hamiltonian of the form $\varepsilon^*\overline{a} + \varepsilon\overline{a}^\dagger$, as opposed to the bare displacement Hamiltonian $\varepsilon^*a + \varepsilon a^\dagger$ of the drive that appears in Eq.~\eqref{eq:H-rf}. This mismatch between bare and dressed states in the drive is the source of the stray population that leaks out of the correct eigenladder during ring-up. To show this mismatch in a simple way, we first focus on the ring-up from an initial ground state $|0,0\rangle=\overline{\ket{0,0}}$. In this case the dominant leakage occurs to the eigenladder $\overline{\ket{n,1}}$, with negligible second-order leakage to the other eigenladders. [As discussed later, the following derivation may be readily generalized to other initial states, such as $\overline{\ket{0,1}}$ in Figs.~\ref{fig:rho-t}(c) and \ref{fig:rho-t}(d).]

Focusing only on the coupling between eigenladders $\overline{\ket{n,0}}$ and $\overline{\ket{n,1}}$, for $n\ll n_c$ we can write \cite{Boissonneault2009,Sete2014}
    \be
    a \approx \bar{a} - \frac{g}{\Delta} \, \overline{\sigma}_- , \,\,\, \Delta = \omega_{\rm r}-\omega_{\rm q}, \,\,\, \overline{\sigma}_- = \sum_n  \overline{|n,0}\rangle\langle\overline{n,1|} ,
    \label{eq:a-expansion}\ee
where $ \overline{\sigma}_-$ is the qubit lowering operator in the eigenbasis. It is natural to guess that at $n\agt n_c$ the resonator-qubit detuning $\Delta$ should change because of the ac Stark shift, and therefore Eq.\ (\ref{eq:a-expansion}) can be replaced with approximation
    \be
    a \approx \bar{a} - \sum_n \frac{g}{\Delta_n} \, \overline{|n,0}\rangle\langle\overline{n,1|} , \,\,\,   \Delta_n = \overline{E}_{n+1,0} - \overline{E}_{n,1},
    \label{eq:a-expansion2}\ee
where $\Delta_n$ is the qubit-resonator detuning with account of the ac Stark shift, $\omega_{\rm q}(n)=\overline{E}_{n,1} - \overline{E}_{n,0}$ (see Appendix of \cite{Sete2015}). We did not prove Eq.\ (\ref{eq:a-expansion2}) analytically, but we checked numerically that this approximation works well, at least for our range of parameters. Additionally approximating $\Delta_n \approx \Delta_{\bar{n}}$ for a dressed coherent state with $\bar{n}=|\alpha|^2$, from Eq.\ (\ref{eq:a-expansion2}) we obtain \be
    a \approx \bar{a} -  \frac{g}{\Delta_{\bar{n}}}\, \overline{\sigma}_-.
    \ee
(For non-integer $\bar{n}$, we can use the nearest integer or the more precise method of averaging $\Delta_n$ over the state.) Note that for a constant resonant drive, the average number of photons increases as $\bar{n}(t)\approx |\varepsilon t|^2$, before the changing resonator frequency (\ref{eq:eigenfreq}) starts affecting the resonance.

Thus, the drive term in the Hamiltonian can be approximately expanded in the eigenbasis as
 \be \label{eq:pump-eigen}
\varepsilon^* a + \varepsilon a^\dagger \approx (\varepsilon^* \bar{a} + \varepsilon \bar{a}^\dagger) - \frac{g}{\Delta_{\bar{n}}}
\left( \varepsilon^*\overline{\sigma}_- + \varepsilon\overline{\sigma}_+ \right),
  \ee
where  $\overline{\sigma}_+ = (\overline{\sigma}_-)^\dagger$.  The first term of this effective drive produces dressed coherent states, while the second term couples the lowest two eigenladders to cause leakage.

The coupling essentially ``copies'' the dressed coherent state from the correct eigenladder $\overline{|n,0\rangle}$ to the neighboring eigenladder. The resulting copy has a relatively small magnitude because $g/\Delta_{\bar{n}} \ll 1$ and also because the two eigenladders have a significant frequency shift due to differing energies. Thus, we assume approximately the same dressed coherent state $\alpha(t)$ in both eigenladders and use the joint state of the form $\ket{\psi} \approx \ket{\alpha(t)}_0 + c(t)\,\ket{\alpha(t)}_1$, where the small amplitude $c(t)$ quantifies the leakage to the $\overline{\ket{n,1}}$ eigenladder, so that the stray population is  $P_\mathrm{stray} = |c|^2 \ll 1$. In this case we can approximately write $c = \bra{\psi}\overline{\sigma}_-\ket{\psi}$, and thus find the evolution $\dot{c} = \bra{\psi}\,i\left[H_\mathrm{rot},\,\overline{\sigma}_-\right]\,\ket{\psi}$, which simplifies to
\begin{align}\label{eq:c-dot}
	\dot{c} & \approx i\, \frac{\varepsilon g}{\Delta_{\bar{n}}} + i\Omega_{\bar{n}}\,c,
\end{align}
where $\Omega_{\bar{n}} = \Delta_{\bar{n}}+ \omega_\mathrm{d} - \omega_{\rm r}$ is the oscillation frequency (note that $\Omega_{\bar{n}} = \Delta_{\bar{n}}$ for a resonant drive). The steady state for this evolution (assuming a slowly changing $\bar{n}$),  $\dot{c}_\mathrm{s.s.} = 0$, corresponds to the steady-state leakage population
\begin{equation}\label{eq:pstray}
  P_\mathrm{s.s.} = |c_\mathrm{s.s.}|^2 = \left|\frac{\varepsilon g}{\Omega_{\bar{n}}\Delta_{\bar{n}}}\right|^2 .
\end{equation}

For a drive that is suddenly turned on, as in Fig.~\ref{fig:rho-t}(a), the stray population will oscillate to reach a maximum
\begin{equation}\label{eq:pmax}
  P_\mathrm{max} = |2c_\mathrm{s.s.}(0)|^2 = 4P_\mathrm{s.s.}(0)=4  \left|\frac{\varepsilon g}{\Omega_0\Delta}\right|^2,
\end{equation}
which is close to the numerical value for $P_\mathrm{max}$ in Fig.~\ref{fig:rho-t}(a).
As discussed below, the oscillations eventually dephase, so we would expect the value $P_{\rm stray}=P_{\rm max}/2$ after that. However, by the time it occurs, $P_{\rm s.s.}$ in Eq.\ (\ref{eq:pstray}), shown by the dashed black line in Fig.~\ref{fig:rho-t}(a), significantly decreases because $\bar{n}$ is already large. As a result, we  expect the value
    $P_{\rm stray}=P_{\rm s.s.}(0) +  P_\mathrm{s.s.} (t)$
after decay of the oscillations. (Here the first term comes from continuing dephased oscillations while the second term comes from the moving center of oscillations on the complex plane of $c$.) This formula is also close to the numerical result in Fig.~\ref{fig:rho-t}(a).

\begin{figure*}[t]
	\begin{center}
\includegraphics[width=\textwidth]{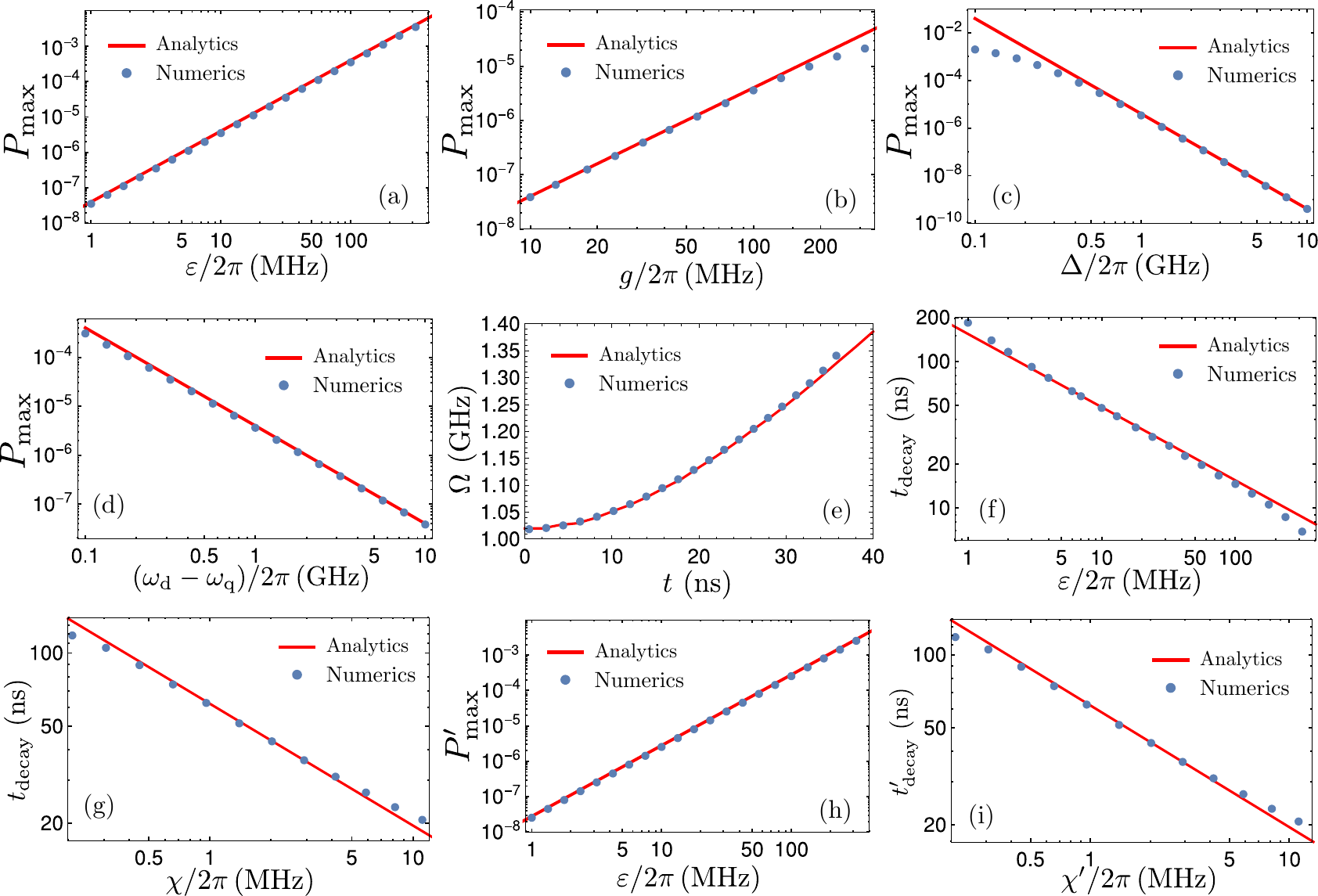}
    \end{center}
    \caption{Model validation for stray population $P_\mathrm{stray}$ in the neighboring eigenladder, using a sudden resonant [off-resonant in (d)] drive and starting with $|0,0\rangle$ (a-g) or $\overline{|0,1\rangle}$ (h,i). Panels (a--d): Testing of Eq.\ (\ref{eq:pmax}) for the maximum stray population $P_\mathrm{max}$ against numerical results, by varying (a) the drive amplitude $\varepsilon$, (b) coupling $g$, (c) resonator-qubit detuning $\Delta$, and (d) drive frequency $\omega_{\rm d}$. (e): Testing that the time-dependent oscillation frequency evolves as $\Omega_{\bar{n}}=\Delta_{\bar{n}}$  given by Eq.\ (\ref{eq:a-expansion2}). (f,g):
    Testing of Eq.\ (\ref{eq:t_decay}) for the decay time $t_{\rm decay}$ of the eigenladder oscillations [as in Fig.~\ref{fig:rho-t}(a)], using a prefactor of 1.23 for decay to $1/3$ amplitude.
   (h,i): Similar to panels (a,g), but for the leakage to the second excited eigenladder $\overline{|n,2\rangle}$ starting from the excited state $\overline{|0,1\rangle}$; in this case Eqs.\ (\ref{eq:pmax}) and (\ref{eq:t_decay}) need the following replacements: $g\mapsto\sqrt{2}\,g$, $\Delta \mapsto \Delta + \eta$, $\Omega_0 \mapsto \Omega_0 + \eta$, $\chi\mapsto \chi' = \omega_{\rm r}^{(2)} -\omega_{\rm r}^{(1)}$. In all panels blue dots show numerical results, while red lines are calculated analytically. We use the following parameters: $\omega_\mathrm{r}/2\pi=6\,\text{GHz}$, $\omega_\mathrm{q}/2\pi=5\,\text{GHz}$, $\eta/2\pi=200\,\text{MHz}$, $g/2\pi=100\,\text{MHz}$, $\varepsilon/2\pi=10\,\text{MHz}$, except for parameters, which are varied, and in (g) $\varepsilon/2\pi=50\,\text{MHz}$ and in (h,i) $\eta/2\pi=300\,\text{MHz}$.
   } \label{fig:eigen-num-model}
\end{figure*}

Figures \ref{fig:eigen-num-model}(a--d) show in more detail that the functional form of Eq.~\eqref{eq:pmax} agrees well with the numerically obtained maximum stray populations $P_{\rm max}$ in the case of a sudden drive. In contrast, when the drive $\varepsilon(t)$ is adiabatically increased from zero, then the stray population closely follows the time-dependent steady state $P_\mathrm{s.s.}$ of Eq.~\eqref{eq:pstray}, as shown in Fig.~\ref{fig:rho-t}(b). Our analysis based on Eq.\ (\ref{eq:c-dot}) predicts that in the diabatic case of a sudden drive, the oscillation frequency $\Omega_{\bar{n}}$ should increase when $\bar{n}$ increases. This is checked in Fig.\ \ref{fig:eigen-num-model}(e); agreement with numerical results is again very good.

Now let us discuss the decay of oscillations seen in Fig.~\ref{fig:rho-t}(a), which is somewhat surprising since our model does not include any decoherence. Numerical results show that the oscillations decay only for a resonant drive (for a strongly off-resonant drive, $\bar{n}\ll 1$ and oscillations do not decay). Therefore, we assume a resonant drive, so that $\bar{n}(t)\approx |\varepsilon t|^2$. Let us now take into account the spread in photon number,  $\bar{n} \pm \sqrt{\bar{n}}$, which produces a corresponding spread in oscillation frequency $\Omega_{n}=\Delta_n$ in Eq.\ (\ref{eq:c-dot}) that dephases the oscillations.
At sufficiently low photon number (up to several $n_c$), we can use the approximation
\be\label{eq:approx-Delta}
  \Delta_n \approx \Delta - 2\chi n, \,\,\, \chi \approx -\frac{\omega_{\rm r}}{\omega_{\rm q}} \, \frac{g^2\eta}{\Delta (\Delta +\eta)},
\ee
which produces the spread of oscillation frequency in Eq.\ (\ref{eq:c-dot}) with the standard deviation
$  \delta \Omega \simeq 2\chi \sqrt{\bar{n}}\approx 2\chi |\varepsilon| t$.
This implies that the corresponding accumulated phase difference after a time $t$ is
$\delta \varphi = \int_0^t \, \delta\Omega\, dt' \approx \chi|\varepsilon| t^2$. Assuming that a phase accumulation of $|\delta\varphi| \simeq 1$ indicates a significant level of dephasing, this estimate yields an oscillation decay time
    \begin{equation}\label{eq:t_decay}
	t_\mathrm{decay} \simeq |\chi \varepsilon |^{-1/2},
    \end{equation}
with an unknown prefactor on the order of 1. This estimate crudely agrees with the oscillation decay in Fig.~\ref{fig:rho-t}(a). For a more detailed analysis we checked the numerical dependence of the decay time on $\varepsilon$ and $\chi$ in  Figs.~\ref{fig:eigen-num-model}(f) and \ref{fig:eigen-num-model}(g). The agreement is quite good using a prefactor of 1.23 in Eq.\ (\ref{eq:t_decay}), when the decay time is defined numerically as decay of the probability oscillations [as in Fig.~\ref{fig:rho-t}(a)] to $1/3$ of initial amplitude. Note that this derivation predicts a crudely Gaussian envelope of oscillation decay for $\sqrt{P_{\rm stray}(t)}$, and this prediction also agrees with the numerical results (though not quite well because of the change of the oscillation center $c_{\rm s.s.}$ over time).

Simple modifications of the above derivation are sufficient to describe the stray populations when starting from a different initial state. As an example, let us consider an initially excited qubit state $\overline{\ket{0,1}}$. In this case there will be two neighboring eigenladders that interact: the ground eigenladder $\overline{\ket{n,0}}$, and the second excited eigenladder $\overline{\ket{n,2}}$. Stray population that leaks to the ground eigenladder will oscillate precisely as before between the ground and excited eigenladders, reproducing Eqs.~\eqref{eq:pstray},  \eqref{eq:pmax},  and \eqref{eq:t_decay}; this equivalence due to symmetry is emphasized in Fig.~\ref{fig:rho-t}(c). In contrast, the stray population leaking to the second excited eigenladder $\overline{\ket{n,2}}$ oscillates between excited and second-excited eigenladders, so behaves somewhat differently. We modify our derivation starting from Eq.~\eqref{eq:pump-eigen} to include only the interaction between the eigenladders $\overline{\ket{n,1}}$ and $\overline{\ket{n,2}}$, which yields the following parameter replacements: $g \to \sqrt{2}\,g$, $\Delta_n \to \overline{E}_{n+1,1}-\overline{E}_{n,2}$, $\Delta \to \Delta + \eta$, $\Omega_0 \to \Omega_0 + \eta$, and $2\chi \to 2\chi' = \omega_{\rm r}^{(2)}(0) - \omega_{\rm r}^{(1)}(0)$. Thus, the equivalents of Eqs.~\eqref{eq:pstray} and \eqref{eq:t_decay} at low $n$ are
\begin{eqnarray}
  && P_\mathrm{s.s.}' = \left|\frac{\sqrt{2}\,\varepsilon g}{(\Delta + \eta - 2\chi'\bar{n})(\Omega_0 + \eta - 2\chi'\bar{n})}\right|^2, \\
  && t_\mathrm{decay}' \simeq  |\chi'\varepsilon |^{-1/2}.
\end{eqnarray}
These equations agree with the numerical results shown in Fig.~\ref{fig:rho-t}(d) and Fig.~\ref{fig:eigen-num-model}(h,i).

Our analysis shows that the stray population of an ``incorrect'' eigenladder considered in this section should be quite small for typical experimental parameters. The case of an adiabatically increased drive is more experimentally relevant, so let us use Eq.\ (\ref{eq:pstray}) and crudely estimate the effect as $P_{\rm stray}\sim (\varepsilon g/\Delta^2)^2$. Then for $g/2\pi \simeq 100$ MHz, $\Delta/2\pi \simeq 1$ GHz, and $\varepsilon/2\pi \simeq 50$ MHz (such drive pumps $\sim 10$ photons within first 10 ns), we obtain $P_{\rm stray}\sim 3\times 10^{-5}$. Even if $\Delta/2\pi$ is decreased to 500 MHz in this estimate and $\varepsilon/2\pi$ is increased to 100 MHz (40 photons within first 10 ns), the resulting value $P_{\rm stray}\sim 2\times 10^{-3}$ still remains quite small. Therefore, this should not significantly affect the qubit measurement error, at least for present-day experiments. (Recall that we neglected qubit energy relaxation, dressed dephasing, Purcell relaxation, and non-RWA effects, which can be responsible for much larger population transfer to incorrect eigenladders.)

\subsection{Infidelity from shearing}\label{sec:shearing}

The second contribution to the infidelity of the dressed coherent state approximation in Eq.~\eqref{eq:fid} is due to infidelity $1-F_{\rm c}$ within the correct eigenladder. As seen in Fig.~\ref{fig:infid}, it becomes increasingly important at longer evolution times, when the number of photons $\bar{n}$ becomes large. As discussed below, this infidelity arises from the effective nonlinearity of the resonator due to its interaction with the transmon. This nonlinearity
produces a shearing effect on the evolution of the dressed coherent state that squeezes the state.

\begin{figure*}[t]
	\begin{center}
		\includegraphics[width=0.75\textwidth]{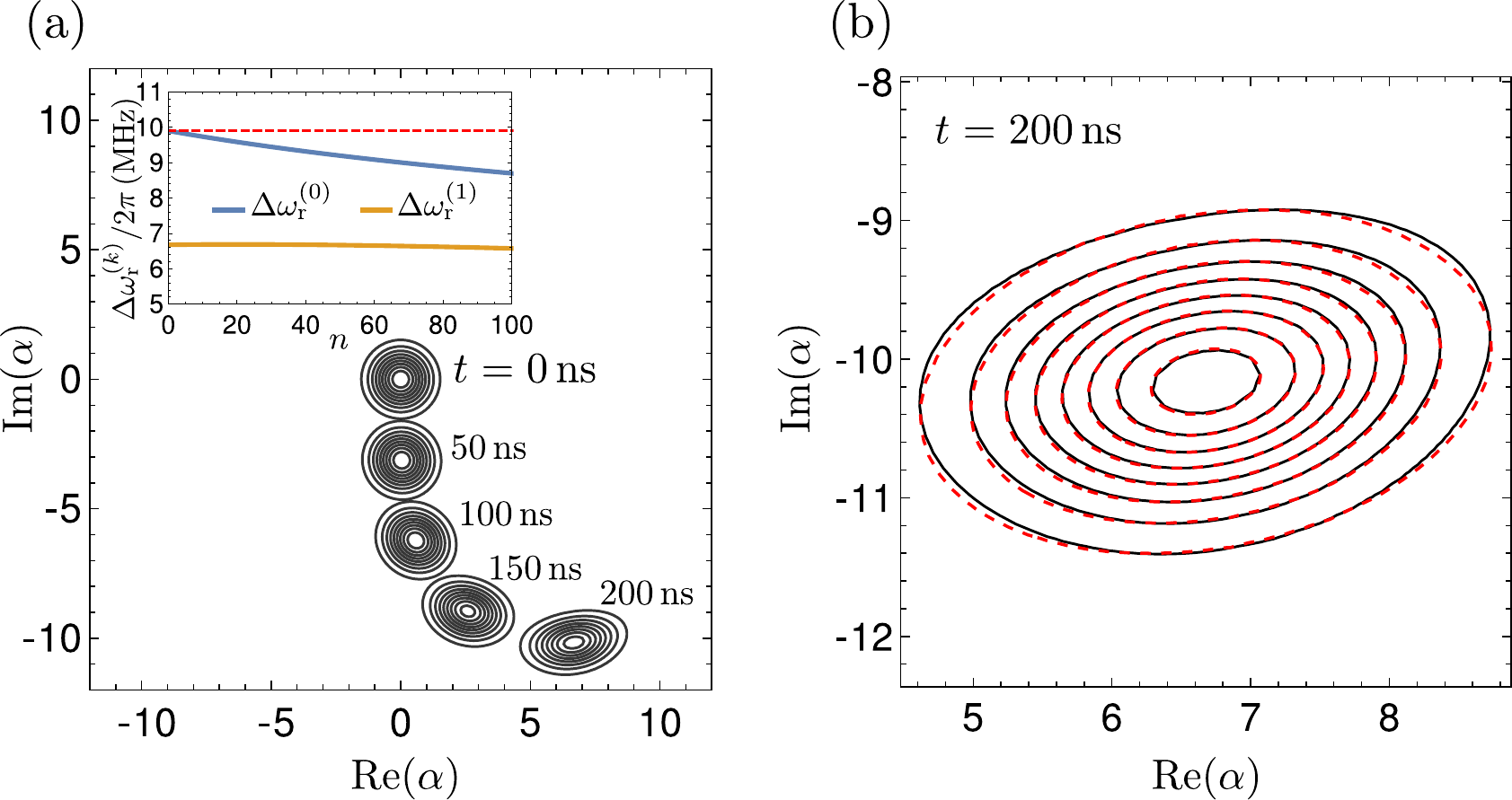}
    \end{center}
    \caption{(a) Numerically simulated evolution of the dressed Husimi Q-function for the state remaining in the correct eigenladder, given an initial state of $\ket{0,0}$ and a resonant drive. Snapshots taken at $50\, \text{ns}$ intervals show the progressive shearing of the state caused by resonator nonlinearity. Inset: $n$-dependence of the difference $\Delta\omega_{\rm r}^{(k)}=\omega_{\rm r}^{(k)}-\omega_{\rm r}$ between the effective and bare resonator frequencies. The solid blue (upper) line shows $\Delta\omega_{\rm r}^{(0)}(n)$ for the ground-state eigenladder, the solid orange (lower) line shows $\Delta\omega_{\rm r}^{(1)}(n)$ for the excited-state eigenladder, and the red dashed line indicates the applied drive frequency.
     (b) Detail of the Q-function at $200\, \text{ns}$. The analytical result for a \emph{dressed squeezed state} (dashed red) shows good agreement with the numerically simulated state (solid black). The agreement is significantly better for earlier times (not shown).  Parameters are: $\omega_\mathrm{r}/2\pi=6\,\mathrm{GHz}$, $\omega_\mathrm{q}/2\pi=5\,\mathrm{GHz}$, $\eta/2\pi=200\,\mathrm{MHz}$, $g/2\pi=100\,\mathrm{MHz}$, and $\varepsilon/2\pi=10\,\mathrm{MHz}$. The contours of the Q-function are drawn at the levels of $0.1/\pi,0.2/\pi,\dots 0.8/\pi$.
    } \label{fig:squeeze}
\end{figure*}

Numerically, this distortion is clearly seen by plotting the Husimi Q-function of the renormalized state $\ket{\psi}_k$ [defined as in Eq.~\eqref{eq:statesplit}] that remains within the correct eigenladder,
\begin{equation}\label{eq:Q}
	Q_\psi(\alpha) = \frac{1}{\pi} \left| {}_k\braket{\alpha}{\psi}_k \right|^2,
\end{equation}
where $\ket{\alpha}_k$ is a dressed coherent state as in Eq.~\eqref{eq:DCS}. The contour plots of $Q_\psi(\alpha)$ in the complex plane of $\alpha$ are shown in Fig.~\ref{fig:squeeze}(a) for a numerically simulated ring-up evolution, starting with the state $|0,0\rangle$ (there are five snapshots at time moments separated by 50 ns).
If the state $\ket{\psi}_k$ were a perfect dressed coherent state $\ket{\psi}_k = \ket{\beta}_k$ centered at $\beta = {}_k\bra{\psi} \bar{a} \ket{\psi}_k$, it would have a Q-function $Q_\psi(\alpha) = e^{-|\alpha-\beta|^2}/\pi$ with circular contours. However, Fig.~\ref{fig:squeeze}(a) clearly shows a progressive distortion of the initial circular profile into a squeezed ellipse as the average photon number increases. We will prove later that $|\psi\rangle_k$ is indeed a close approximation of a (minimum-uncertainty) squeezed state in the eigenbasis $\overline{|n,k\rangle}$ -- see Fig.\ \ref{fig:squeeze}(b).

The squeezing distortion in Fig.~\ref{fig:squeeze} is similar to the self-developing quadrature squeezing discussed in Ref.\ \cite{Sete2013} for the catch-disperse-release measurement protocol (e.g., compare Fig.~\ref{fig:squeeze} with the figures in the Supplemental Material of \cite{Sete2013}). In that protocol, the squeezing was shown to significantly decrease the measurement error. In general, the self-developing squeezing can either increase or decrease the measurement error depending on the angle of the squeezing axis, and the analysis is clearly important for practical qubit measurements.
A strong self-developing squeezing has been observed experimentally in Ref.\ \cite{Kirchmair2013}.

The reason for the self-developing squeezing is the nonlinearity of the transmon, which makes the effective resonator frequency $\omega_{\rm r}^{(k)}(n)$ dependent on the number of photons $n$ -- see Eq.\ (\ref{eq:eigenfreq}) and the inset of Fig.~\ref{fig:squeeze}(a).
Qualitatively, this $n$-dependence causes parts of the circles in Fig.~\ref{fig:squeeze}(a) with different distances $|\alpha|$ from the origin to rotate with slightly different angular velocities, thus shearing the circular profile of an initially coherent state as it evolves. Note that in the case of a constant derivative $d\omega_{\rm r}^{(k)}(n)/dn$, the shearing rate should grow with $|\alpha|$ because $dn = 2|\alpha|\, d|\alpha|$; thus, the effect becomes more important for larger photon numbers. Also note that the drift of the resonator detuning from the drive with $n$ could be compensated for by changing the drive frequency (chirping); however, this does not affect the shearing, since it originates from the frequency variation within the photon number uncertainty $\bar{n}\pm \sqrt{\bar{n}}$.

It is easy to analyze the shearing effect in the absence of the drive. If at $t=0$ we have a dressed coherent state given by Eq.\ (\ref{eq:DCS}) (with notation $\alpha$ replaced by $\beta$), then it obviously evolves as
    \be
    |\psi (t) \rangle_k = e^{-|\beta|^2/2} \sum_n \frac{\beta^n}{\sqrt{n!}} \, e^{-i \overline{E}_{n,k} \, t} \, \overline{|n,k\rangle} ,
    \ee
where the eigenenergies $\overline{E}_{n,k}$ are in the rotating frame $\omega_{\rm d}$, i.e., with subtracted terms $(n+k)\omega_{\rm d}$. Let us expand these energies up to the second order in the vicinity of $\bar{n}=|\beta|^2$ as $\overline{E}_{n,k} \approx \overline{E}_{\bar{n},k}+ \omega_{\rm r}^{(k)}(\bar{n})\, (n-\bar{n})+ \frac{1}{2} (d \omega_{\rm r}^{(k)}/dn)_{|\bar{n}}(n-\bar{n})^2$, where the resonator frequencies $\omega_{\rm r}^{(k)}(n)$ are also in the rotating frame (i.e., with subtracted $\omega_{\rm d}$) and we neglect discreteness of $n$ by assuming $\bar{n}\gg 1$ and sufficiently small nonlinearity. This gives
    \begin{eqnarray}
  &&  |\psi (t) \rangle_k \approx e^{-|\beta|^2/2} \sum_n \frac{[\beta(t)]^n}{\sqrt{n!}} \, e^{-i q \,(n-\bar{n})^2} \, \overline{|n,k\rangle} , \qquad
    \label{eq:psi-sheared}\\
&&    \dot{\beta} =  -i\omega_{\rm r}^{(k)}(\bar{n}) \, \beta , \,\,\,\,\,
   \dot{q} = \frac{1}{2} (d \omega_{\rm r}^{(k)}/dn)_{|\bar{n}} ,
    \label{eq:q-dot}\end{eqnarray}
where we neglected the overall phase of $|\psi (t) \rangle_k$. Thus, to leading order in $|n-\bar{n}|$, the effect is an appearance of the quadratic phase factor $e^{-i q \,(n-\bar{n})^2}$ and an obvious rotation of $\beta(t)$ when the effective resonator frequency $\omega_{\rm r}^{(k)}(\bar{n})$ is not exactly on resonance with the drive. The presence of the growing quadratic-term coefficient $q$ in the phase factor leads to a deviation from the dressed coherent state, for which $q=0$. (We restrict our attention to the case $q\ll 1$; very interesting effects beyond this regime, including state revival and formation of ``cat'' states, have been observed in \cite{Kirchmair2013}.)

It is easy to see that the infidelity of the sheared state (\ref{eq:psi-sheared}) compared with the dressed coherent state $|\beta\rangle_k$ is
    \be
      1-F_{\rm c} \approx q^2 \, \overline{(n-\bar{n})^4} \approx 3 (q\, |\beta|^2)^2,
    \ee
assuming $1-F_{\rm c} \ll 1$ and $\bar{n}\gg 1$.
This infidelity grows in time because of the $q$-evolution (\ref{eq:q-dot}) due to the nonlinearity. However, the state evolution due to drive (in a locally linear system) should preserve $1-F_{\rm c}$ because both states ($|\psi\rangle_k$ and $|\beta\rangle_k$) are equally displaced  within the complex plane of $\alpha$ (mathematically, because the standard displacement operator is unitary). Therefore, if the state remains in the form (\ref{eq:psi-sheared}), then
    \be
\frac{d}{dt}(q\,|\beta|^2) =\dot{q} \, |\beta|^2= \frac{\bar{n}}{2} \,  (d\omega_{\rm r}^{(k)}/dn)_{|\bar{n}}.
    \label{eq:dot-q-beta2}\ee
In particular, if $\bar{n}\approx (\varepsilon t)^2$ for a resonant drive and the derivative $d\omega_{\rm r}^{(k)}/dn$ does not significantly depend on $n$ [see inset in Fig.~\ref{fig:squeeze}(a)], then $q\bar{n} \simeq (d\omega_{\rm r}^{(k)}/dn)\, \varepsilon^2 t^3/6$, and the infidelity is
    \be
    1-F_{\rm c} \simeq \frac{1}{12} \, [\varepsilon^2 t^3 \, (d\omega_{\rm r}^{(k)}/dn)]^2 .
    \label{eq:infid-c}\ee

\begin{figure}[t]
	\begin{center}
		\includegraphics[width=8cm]{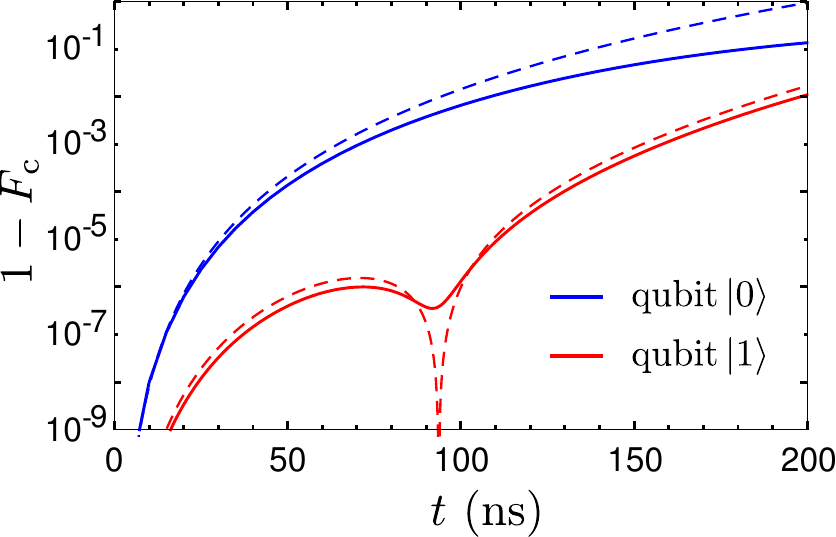}
    \end{center}
    \caption{Infidelity $1-F_{\rm c}$ of the dressed coherent state approximation within the initial eigenladder, starting with $|0,0\rangle$ (upper lines, blue) or $\overline{|0,1\rangle}$ (lower lines, red). Parameters are the same as for Fig. \ref{fig:infid}. The solid lines show numerical results, upper (blue) dashed line is calculated via Eq.\ (\ref{eq:infid-c}) with the frequency derivative taken at $n=0$, and the lower (red) dashed line is calculated by integrating Eq.\ (\ref{eq:dot-q-beta2}).
    } \label{fig:infid-c}
\end{figure}

This is a very crude estimate because $d\omega_{\rm r}^{(k)}/dn$ depends on $n$, the approximation $\bar{n}\approx (\varepsilon t)^2$ works only at small $t$  and, most importantly, the state during the evolution does not remain in the form (\ref{eq:psi-sheared}), as discussed in the next section. [The form (\ref{eq:psi-sheared}) is no longer applicable when the motion of the Q-function center shown in Fig.\ \ref{fig:squeeze}(a) deviates from a straight line.] Nevertheless, comparison with numerical results in Fig.\ \ref{fig:infid-c} shows that Eq.\ (\ref{eq:infid-c}) gives a reasonable  estimate of the infidelity. The blue (upper) solid line in Fig.\ \ref{fig:infid-c} is identical to the orange line in Fig.\ \ref{fig:infid} and shows the numerically calculated $1-F_{\rm c}$ for the evolution  starting with $|0,0\rangle$. The blue (upper) dashed line is obtained using Eq.\ (\ref{eq:infid-c}) with $d\omega_{\rm r}^{(0)}/dn$ calculated at $n=0$. It fits the solid line well at short times, and then deviates up, mostly because $|d\omega_{\rm r}^{(0)}/dn|$ decreases with $n$ [see inset in Fig.\ \ref{fig:squeeze}(a)] while analytics still uses the value at $n=0$. The red (lower) solid line in Fig.\ \ref{fig:infid-c} shows $1-F_{\rm c}$ for the evolution starting with $\overline{|0,1\rangle}$. This  infidelity is crudely two orders of magnitude less than for the blue (upper) line because the derivative $|d\omega_{\rm r}^{(1)}/dn|$ within the excited-state eigenladder is much smaller than that for the ground state [see inset in Fig.\ \ref{fig:squeeze}(a)].  The infidelity $1-F_{\rm c}$ shows a dip near 100 ns. This is because $\omega_{\rm r}^{(1)}(n)$ increases for $n<20$ and  decreases for $n>20$; therefore $q|\beta|^2$ in Eq.\ (\ref{eq:dot-q-beta2}) first increases and then decreases, passing through zero. At the point of passing zero we expect $1-F_{\rm c}=0$, thus producing the dip; numerically it is not zero because the form (\ref{eq:psi-sheared}) is only an approximation. Since $d\omega_{\rm r}^{(1)}/dn$ depends on $n$ very significantly (even changing the sign), we cannot use Eq.\ (\ref{eq:infid-c}), so instead we have integrated Eq.\ (\ref{eq:dot-q-beta2}) to obtain the red (lower) dashed line in Fig.\ \ref{fig:infid-c}. As we see, it agrees well with the solid line. If the integration of Eq.\ (\ref{eq:dot-q-beta2}) is also done for the evolution starting with $|0,0\rangle$, then the result is significantly closer to the blue solid line than the blue dashed line.

Note that the states with a quadratic phase factor as in Eq.\ (\ref{eq:psi-sheared}) have been
discussed in optics long ago \cite{Tanas1984,*Tanas1989,Milburn1986,Kitagawa1986}. It was shown that these states are squeezed in the broad sense that variance of a quadrature operator can be smaller than that for a coherent state. However, to the best of our knowledge, it was never shown that such states with large $\bar{n}$ can be represented as squeezed states in the narrow sense, i.e., they are close to satisfying the minimum-uncertainty condition. Moreover, it was often emphasized that the states described by Eq.\ (\ref{eq:psi-sheared}) are not the minimum-uncertainty states, because for sufficiently large $q$ they have crescent-like shape of the Q-function instead of the elliptical shape, and for even larger $q$ the shape becomes a ring-like one (see experiment \cite{Kirchmair2013}). In contrast, in the next section we will show that in the practically interesting regime these states are quite close to the squeezed states in the narrow sense. This is because for large $\bar{n}$ the squeezing factor is determined by $q|\beta|^2$, while significant deviation from a minimum-uncertainty squeezed state starts at $|q\beta|\agt 0.1$; therefore the squeezing becomes significant already for such values of $q$, for which the deviation (crescent-like shape) is still quite small -- see Fig.~\ref{fig:squeeze}(b). In the next section we will also derive simple evolution equations for these squeezed states.

\section{Dressed squeezed state model}\label{sec:squeezed}

As discussed in the previous section, transmon-induced nonlinearity of the resonator (i.e., frequency dependence on the number of photons) evolves a dressed coherent state into a sheared state of the form (\ref{eq:psi-sheared}) with quadratic phase factor. Unfortunately, it is not easy to describe evolution of this sheared state due to drive. In some sense this is because an evolution due to drive is naturally described in the phase space (which is almost always used in optics), while the sheared state representation requires Fock space. We will be able to solve this dilemma by showing that the sheared state (\ref{eq:psi-sheared}) is actually close to a (minimum-uncertainty) squeezed state in the eigenbasis, which we call a dressed squeezed state. Evolution of a squeezed state due to drive can be easily described in the phase space, while its evolution due to nonlinearity can be easily described in the Fock space. Thus, if we have a reasonably simple conversion between the Fock and phase spaces for squeezed states, we can describe the state evolution due to both nonlinearity and drive. This simple conversion is possible only for large $\bar{n}$, which is an important assumption for our derivation below (in practice, it is still well applicable for the dynamics starting with the vacuum state).

\subsection{Dressed sheared Gaussian state}

In this section we prove that for sufficiently large number of photons, the (dressed) sheared  state is approximately equivalent to a (dressed) minimum-uncertainty squeezed state.

For $|\beta|^2\gg 1$  we can use a Gaussian approximation for the wavefunction (\ref{eq:psi-sheared}) in the Fock space. Let us introduce a more general (dressed) sheared Gaussian state as
    \begin{eqnarray}
    && \hspace{-0.1cm} |\beta , K, W\rangle_k = \sum_n  \frac{1}{(2\pi W |\beta|^2)^{1/4}} \, \exp \left[ -\frac{(n-|\beta|^2)^2}{4W|\beta|^2} \right]
    \nonumber \\
    && \hspace{0.9cm} \times \exp[ i n\, {\rm arg} (\beta)] \, \exp\left[ -i\frac{ K(n-|\beta|^2)}{|\beta|^2}\right] \overline{|n,k\rangle}\, , \qquad
    \label{eq:sheared-Gaussian}\end{eqnarray}
in which we used the new notation $K=q|\beta|^2$ and also introduced a new parameter $W=\sigma_{n}^2/\sigma_{n,\rm cs}^2$, which is the variance $\sigma_n^2$ of the Gaussian $n$-distribution compared with the variance $\sigma_{n,\rm cs}^2=|\beta|^2$ of a dressed coherent state, so that $w=\sqrt{W}$ is the relative width of the $n$-distribution. Thus, the sheared Gaussian state is characterized by 4 parameters: $\beta$ has the standard optical meaning, $K$ characterizes the shearing, $W$ characterizes the relative width of photon number distribution, and $k$ labels the eigenladder. We assume that $K$ and $W$ are on the order of 1, while $|\beta|^2 \gg 1$. Note that the term $i n\, {\rm arg}(\beta)$ can be replaced with $i(n-|\beta|^2)\, {\rm arg}(\beta)$; this changes only the unimportant overall phase of the state, but clarifies the role of ${\rm arg}(\beta)$ as the linear-order part of the phase expansion in $n$ around the mean $|\beta|^2$.

We call the form (\ref{eq:sheared-Gaussian}) a hybrid phase-Fock representation, because $\beta$ is borrowed from optical phase space, while $K$ and $W$ are the Fock-space parameters. Note that the state (\ref{eq:sheared-Gaussian}) is not exactly normalized, but the difference from perfect normalization is less than $10^{-5}$ if $|\beta|^2 >\max (20\,W,1/W)$. With a similar accuracy, $\bar{n}=|\beta|^2$ for the average number of photons (excitations in the eigenladder).

The average value of the dressed lowering operator for the state (\ref{eq:sheared-Gaussian}) is
    \be
    \langle \bar{a}\rangle \approx \beta + \frac{2-W-1/W}{8\beta^*} - i\frac{KW}{\beta^*} -\frac{2 K^2 W}{\beta^*} \approx \beta ,
    \label{<a>}\ee
where in the second equality we neglected the terms scaling as $|\beta|^{-1}$. Similarly, neglecting $|\beta|^{-1}$-terms, we find
 \begin{eqnarray}
  \langle \bar{a}^2\rangle \approx \beta^2 +\frac{\beta^2}{|\beta|^2} \, \bigg( \frac{1}{2} -\frac{1}{2W} -4iKW -8 K^2W\bigg). \qquad
 \label{<a2>}\end{eqnarray}
Now let us define (dressed) quadrature operators,
    \be
    X_\varphi = \frac{1}{2}\left(e^{-i\varphi} \, \bar{a}+e^{i\varphi}\, \bar{a}^\dagger  \right), \ee
for which $\varphi$ is the quadrature angle (note that notation $\varphi$ was briefly used for a different quantity in Sec.\ \ref{sec:eigen}). Using Eqs.\ (\ref{<a>}) and (\ref{<a2>}) we find the variance $\sigma_{X\varphi}^2=  \langle X_\varphi^2\rangle -\langle X_\varphi \rangle^2$,
    \begin{eqnarray}
    && \sigma_{X\varphi}^2 = \frac{W+1/W}{8} +2K^2W  +\, KW \sin [2\,{\rm arg}(\beta)-2\varphi ]
        \nonumber \\
     && \hspace{0.9cm}  + \bigg( \frac{W-1/W}{8}-2K^2W\bigg) \cos [2\,{\rm arg}(\beta)-2\varphi ] . \qquad
         \end{eqnarray}

It is easy to check that the $\varphi$-dependence of this variance is exactly what would be expected for a minimum-uncertainty squeezed state. In particular, the product of the minimum and maximum values of $\sigma_{X\varphi}^2$ is the same as for a coherent state,
    \be
    \sigma_{X\varphi, \rm min}^2    \sigma_{X\varphi, \rm max}^2 = 1/16,
    \label{eq:quad-prod}\ee
with
    \begin{eqnarray}
     && \sigma_{X\varphi, \rm min}^2 = \left[ 1+S -\sqrt{(1+S)^2 -1} \right] /4,
     \label{eq:Xphi-min}\\
     && S= 8 K^2 W +(W+1/W-2)/2,
    \label{eq:S-def}\end{eqnarray}
and $\sigma_{X\varphi, \rm max}^2 =[ 1+S +\sqrt{(1+S)^2 -1}]/4$. We see that the degree of squeezing is determined by the parameter $S$, so that $S=0$ corresponds to a (dressed) coherent state. The minimum quadrature variance $ \sigma_{X\varphi, \rm min}^2$ is achieved at the angle $\varphi_{\rm min}=\theta /2$, where
    \begin{eqnarray}
        && \theta = 2 \, {\rm arg}(\beta) + {\rm arctan}\left(\frac{8KW}{16K^2W-W+1/W}  \right)
        \nonumber \\
     && \hspace{1cm} +\,\frac{\pi}{2} \, [1-{\rm sgn}(16K^2W-W+1/W)],
    \label{eq:theta}\end{eqnarray}
and the factor of 2 between $\theta$ and $\varphi_{\rm min}$ is to conform with the standard optical definition of the squeezing parameter discussed later.

Thus, we have proven that for sufficiently large $|\beta |^2$ the (dressed) sheared Gaussian state (\ref{eq:sheared-Gaussian}) is close to a (dressed) minimum-uncertainty squeezed state (despite this is not true for small $|\beta|^2$ \cite{Milburn1986,Kitagawa1986,Kirchmair2013}).
Note that the ``conservation of area'' criterion (\ref{eq:quad-prod}) for a minimum-uncertainty squeezed state is valid for quadratures, but is not valid for the Husimi Q-function shown in Fig.\ \ref{fig:squeeze}, because the Q-function involves convolution with a coherent state, and therefore the width of the short axis can be at most a factor of $\sqrt{2}$ shorter than that of a coherent state.

\subsection{Conversion into squeezed state notations}

Using the standard optical definition \cite{Yuen1976,Gerry2005-book}, a dressed squeezed state should be defined as
\be\label{eq:squeezed}
  \ket{\beta,\xi}_k = \exp[\beta\bar{a}^\dagger - \beta^*\bar{a}]\, \exp \left[\xi^*\frac{\bar{a}^2}{2}-\xi\frac{\bar{a}^\dagger}{2}\right]
  \overline{\ket{0,k}} ,
\ee
where $\xi \equiv r e^{i\theta}$ is the squeezing parameter, while $\beta$ is a displacement in the phase space. The smallest standard deviation $\sigma_{X\varphi, \rm min}$ for the quadrature $X_\varphi$ should then be achieved \cite{Yuen1976,Gerry2005-book} at the angle $\varphi_{\rm min}=\theta/2$ [thus corresponding to our notation in Eq.\ (\ref{eq:theta})], and its value should be $\sigma_{X\varphi, \rm min}=e^{-r}\sigma_{X\varphi, \rm cs}$ compared with the standard deviation $\sigma_{X\varphi, \rm cs}$ for a coherent state. The longest axis is $\sigma_{X\varphi, \rm max}=e^{r}\sigma_{X\varphi, \rm cs}$ at the angle $\varphi_{\rm max}=\theta/2 \pm \pi/2$.

Comparing these standard optical definitions with our approximate results (\ref{eq:quad-prod}), (\ref{eq:Xphi-min}), and (\ref{eq:theta}) for large $|\beta|^2$, we obtain the conversion
    \be
    r = \frac{1}{2}\,\mathrm{arccosh}(S+1),
    \label{eq:r-conversion}\ee
where $S$ is given by Eq.\ (\ref{eq:S-def}), while $\theta$ is given by Eq.\ (\ref{eq:theta}).

It is easy to check that the case $K=0$, $W=1$ corresponds to the dressed coherent state, $\xi=0$. In the absence of shearing, $K=0$, we have a dressed amplitude-squeezed state for $W<1$ [as is obvious from Eq.\ (\ref{eq:sheared-Gaussian})] and a dressed phase-squeezed state for $W>1$ -- see Eq.\ (\ref{eq:theta}), from which $\theta/2={\rm arg}(\beta)$ for $W<1$ and $\theta/2={\rm arg}(\beta)\pm \pi/2$ for $W>1$. As shown in the Appendix, the dressed squeezed state is practically unentangled for large $|\beta|^2$, in spite of a significant entanglement of the qubit-resonator eigenstates.

Using Eqs.\ (\ref{eq:theta}) and (\ref{eq:r-conversion}) we can convert a sheared Gaussian state (\ref{eq:sheared-Gaussian}) with sufficiently large $|\beta|^2$ into a (minimum-uncertainty) squeezed state (\ref{eq:squeezed}). Similarly, we can convert any (minimum-uncertainty) squeezed state with sufficiently large $|\beta|^2$ into a sheared Gaussian state. Most importantly, we know that a squeezed state is simply displaced in the phase space by an action of a drive $\varepsilon(t)$. This means that a sheared Gaussian state (\ref{eq:sheared-Gaussian}) remains a sheared Gaussian state under an action of the drive (assuming large $|\beta|^2$). Since it also keeps the form (\ref{eq:sheared-Gaussian}) under the evolution due to nonlinearity, this form is always preserved (approximately), and therefore it is sufficient for us to characterize the evolution of the state by evolution of only three parameters: $\beta$, $K$, and $W$. We emphasize that this simplicity is possible only for large $|\beta|^2$ or, in other words, for a sufficiently small nonlinearity. In general, the simultaneous evolution due to nonlinearity and drive creates states that cannot be described as (minimum-uncertainty) squeezed states or sheared states. Nevertheless, this approximation works quite well for our system.

 \subsection{Phase-Fock-space evolution of dressed squeezed state}

Now let us derive evolution equations for the parameters $K$, $W$, and $\beta$ of the dressed sheared/squeezed state. We will first consider the evolution in the absence of the drive, then the evolution only due to the drive, and then add up the terms from these evolutions.

Evolution of the dressed sheared state (\ref{eq:sheared-Gaussian}) due to nonlinearity of the resonator is given by Eq.\ (\ref{eq:q-dot}), which leads to
    \be
    \dot{K}=  \frac{1}{2}\, |\beta|^2 (d\omega_{\rm r}^{(k)}/dn)\Big|_{ n=|\beta|^2}.
    \ee
Note that we do not need to take a derivative of $|\beta|^2$ because this type of evolution does not change $|\beta|^2$. In the absence of the drive, the parameter $\beta$ evolves only due to the resonator frequency detuning from the rotating frame,
    \be
\dot{\beta} =  -i\omega_{\rm r}^{(k)}(n)\Big|_{n=|\beta|^2} \beta.
    \label{eq:beta-dot-1}\ee

To derive formulas for the evolution of $\beta$, $K$, and $W$ due to drive $\varepsilon (t)$, we use the fact \cite{Gerry2005-book} that for a squeezed state (\ref{eq:squeezed}) the parameter $\xi$ remains constant, while $\beta$ changes as $\dot{\beta}=-i\varepsilon$. Therefore, the parameters $S$ and $\theta$ given by Eqs.\ (\ref{eq:S-def}) and (\ref{eq:theta}) should remain constant with changing $\beta$. The corresponding evolution $\dot{K}$ and $\dot{W}$ can be found from the system of equations
    \be
    \frac{\partial S}{\partial K} \, \dot{K} +   \frac{\partial S}{\partial W} \, \dot{W}=0,  \,\,\,
    \frac{\partial \theta }{\partial K} \, \dot{K} +   \frac{\partial \theta}{\partial W} \, \dot{W}+ \frac{\partial \theta}{\partial \beta} \, \dot{\beta}=0,
    \label{eq:partial0}\ee
which has the following solution:
    \be
    \dot{W}=8 KW {\rm Re}(\varepsilon /\beta), \,\,\, \dot{K}= \bigg(\frac{1-W^2}{4W^2}-4K^2 \bigg) {\rm Re}(\varepsilon /\beta),
    \ee
where we took into account the equation $\dot{\beta}=-i\varepsilon$. Note that here we should not include evolution of $\beta$ due to detuning, Eq.\ (\ref{eq:beta-dot-1}), because otherwise the angle $\theta$ would not be constant. Also note that in the term $(\partial\theta/\partial\beta)\dot{\beta}$ in Eq.\ (\ref{eq:partial0}) we imply derivatives for both ${\rm Re}(\beta)$ and ${\rm Im}(\beta)$.

Combining the evolution equations both in the absence of a drive and from the drive itself, we finally obtain
    \begin{eqnarray}
     && \dot{W}=8 KW \, {\rm Re}(\varepsilon /\beta),
     \label{eq:evol-W}\\
     && \dot{K}= \bigg(\frac{1-W^2}{4W^2}-4K^2 \bigg) \, {\rm Re}(\varepsilon /\beta)
        \nonumber \\
     &&\hspace{0.8cm} +
     \frac{1}{2}\, |\beta|^2 \, (d\omega_{\rm r}^{(k)}/dn)\Big|_{n=|\beta|^2}\, ,
     \label{eq:evol-K}
     \\
     && \dot{\beta} =  -i\omega_{\rm r}^{(k)}(n)\Big|_{n=|\beta|^2} \beta -i\varepsilon .
    \label{eq:evol-beta}\end{eqnarray}
These equations together with the conversion formulas (\ref{eq:theta}) and (\ref{eq:r-conversion}) is our {\it main result} for the evolution of the dressed squeezed state. They allow very efficient simulation, since they avoid the large dimensionality of the pure Fock-space evolution specified by Eq.~\eqref{eq:H-rf}.
Equations (\ref{eq:evol-W})--(\ref{eq:evol-beta}) are a hybrid between the Fock-space and the phase-space representations, capable of describing evolution of the dressed squeezed state as it rings up due to a coherent drive $\varepsilon$. To our knowledge, this is a novel representation, which was not previously used in optics.

Note that the derivation of these equations assumes large $|\beta|^2$. However, they can be numerically applied even for evolution starting with vacuum, $\beta (0)=0$. [There is no divergence due to the factor of $\beta$ in the denominator, because at small times $\beta =-i\varepsilon t $, and therefore ${\rm Re}(\varepsilon/\beta)={\rm Re}(i/t)=0$.] We used these relatively simple equations to compare with the numerical results for evolution due to Hamiltonian (\ref{eq:H-rf}) in a system with typically $7\times 300$ levels, and found very good agreement. The reason why Eqs.\ (\ref{eq:evol-W})--(\ref{eq:evol-beta}) still work well when starting with the vacuum is that the effect of nonlinearity at short times is small ($K\approx 0$, $W\approx 1$), while by the time when the squeezing due to nonlinearity becomes important, $|\beta|^2$ is already large.
Note, however, that for $|\beta|^2\alt 100$ the sheared and squeezed states are significantly different, and then it is important to use the dressed squeezed state (\ref{eq:squeezed}) [not the sheared state (\ref{eq:sheared-Gaussian})] as the more accurate model for comparison with simulation results.

Figure \ref{fig:squeeze}(b) shows comparison between the Q-function for the numerically calculated state $|\psi\rangle_0$ (solid lines) and for the dressed squeezed state (dashed lines) calculated using Eqs.\ (\ref{eq:evol-W})--(\ref{eq:evol-beta}). [At the end we have converted parameters $K$ and $W$ into the squeezing parameters $r$ and $\theta$ using Eqs.\ (\ref{eq:theta}) and (\ref{eq:r-conversion}), and then calculated the Q-function using the standard formula \cite{Yuen1976} for a squeezed state.] If the parameter $\beta$ is not calculated from Eq.\ (\ref{eq:evol-beta}) but is instead computed as $\beta={}_0\langle \psi |\bar{a}|\psi\rangle_0$, then the visual agreement between the dashed and solid lines becomes insignificantly better. The visible difference between solid and dashed lines is because the numerical state $|\psi\rangle_0$ is not exactly the dressed squeezed state; in particular, for Fig.\ \ref{fig:squeeze}(b) $|q \beta|\sqrt{W}=|K/\beta|\sqrt{W}=0.023$, which is comparable to the value of 0.1, above which a significant crescent-shape appears. The dashed lines in Fig.\ \ref{fig:squeeze}(b) are drawn for the squeezing parameter $r=0.550$. This corresponds to the minimum and maximum quadrature variances of 0.333 and 3.00 compared with the coherent state (0.340 and 3.01 numerically for $|\psi\rangle_0$) and the scaling factors of 0.816 and 1.41 for the short and long axes of the Q-function, compared with the coherent state (numerically 0.81 and 1.43 in the vicinity of the center).

\subsection{Accuracy of dressed squeezed state approximation}

\begin{figure}[t]
	\begin{center}
		\includegraphics[width=\columnwidth]{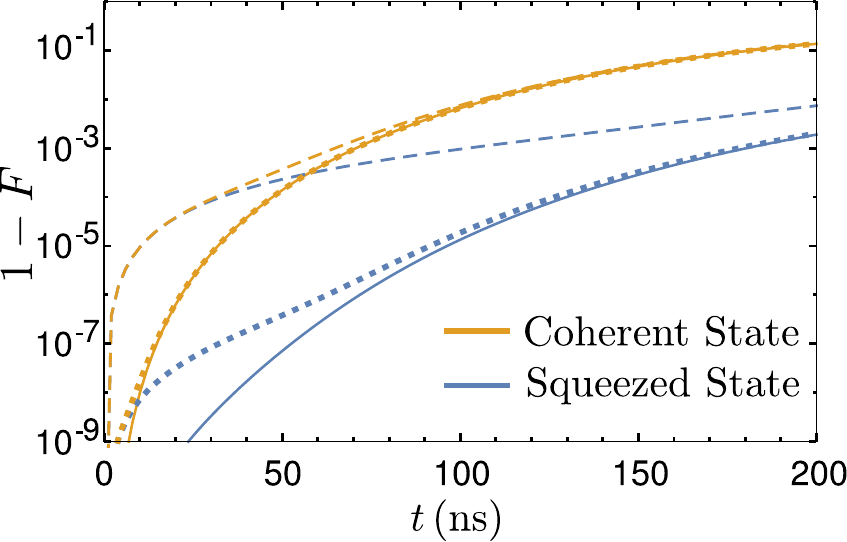}
    \end{center}
\caption{Comparison between the dressed squeezed state and dressed coherent state models within the ``correct'' eigenladder. Parameters are the same as in Figs.~\ref{fig:infid} and  \ref{fig:squeeze}, evolution starts with $|0,0\rangle$. Blue (lower) lines show time dependence of the infidelity $1-F=1-|{}_0\langle \beta,\xi|\psi\rangle_0|^2$ for the dressed squeezed states, orange (upper) lines show infidelity $1-|{}_0\langle \alpha|\psi\rangle_0|^2$ for the dressed coherent states. For solid lines the state centers $\beta(t)$ and $\alpha(t)$ are calculated as average values of the operator $\bar{a}$. For the dashed lines, $\beta(t)$ and $\alpha(t)$ are obtained from Eq.\ (\ref{eq:evol-beta}). For the dotted lines, in Eqs.\ (\ref{eq:evol-W})--(\ref{eq:evol-beta}) we use correction (\ref{eq:tilde-epsilon}) for the drive amplitude. The dressed squeezed state model is about two orders of magnitude more accurate than the dressed coherent state model.
 }\label{fig:DSS}
\end{figure}

To quantify the accuracy of the dressed squeezed state approximation and evolution equations (\ref{eq:evol-W})--(\ref{eq:evol-beta}), we compare the numerically calculated state $|\psi\rangle_0$ for the evolution shown in Fig.\ \ref{fig:infid} (starting with $|0,0\rangle$) with the result from Eqs.\ (\ref{eq:evol-W})--(\ref{eq:evol-beta}) for the sheared Gaussian state, which is then converted into the dressed squeezed state $|\beta,\xi\rangle_0$. The infidelity $1-F= 1-|{}_0\langle \beta,\xi|\psi\rangle_0|^2$ is shown in Fig.\ \ref{fig:DSS} as the dashed blue (lower) line. It can be compared with similar infidelity for the dressed coherent state shown as the dashed orange (upper) line, for which we also used Eq.\ (\ref{eq:evol-beta}). We see that the accuracy of the dressed squeezed state model is much better than for the dressed coherent state model when the infidelity of the latter exceeds $10^{-3}$. However, at short times both infidelities practically coincide and are significantly larger than the coherent-state infidelity $1-F_{\rm c}$ shown in Fig.\ \ref{fig:infid} (also copied as the solid orange line in Fig.\ \ref{fig:DSS}). Since the difference between the orange dashed and orange solid lines is the method of $\alpha(t)$ calculation, either via Eq.\ (\ref{eq:evol-beta}) or as $\alpha = {}_0\langle \psi|\bar{a}|\psi\rangle_0$, this indicates an inaccurate result of Eq.\ (\ref{eq:evol-beta}) for the state center in the phase space. Let us similarly calculate the center of the dressed squeezed state as $\beta = {}_0\langle \psi|\bar{a}|\psi\rangle_0=\alpha$, while the squeezing parameter $\xi$ is still calculated via Eqs.\ (\ref{eq:evol-W})--(\ref{eq:evol-beta}). This produces the blue (lower) solid line in Fig.\ \ref{fig:DSS}, which is crudely two orders of magnitude lower than $1-F_{\rm c}$, thus confirming that the dressed squeezed state approximation is much better than the dressed coherent state approximation.

The reason for the inaccuracy of $\beta(t)$ [or $\alpha(t)$] calculation is rather simple. For the dashed lines in Fig.\ \ref{fig:DSS} we used the bare-basis value for the drive amplitude $\varepsilon$, while within an eigenladder it is actually slightly different. Using properly normalized eigenstates for $n\ll n_c$, it is easy to obtain the second-order correction in Eq.\ (\ref{eq:a-expansion}): $a\approx [1+\frac{1}{2}(g/\Delta)^2\sigma_z] \, \bar{a}-(g/\Delta)\,\overline{\sigma}_-$  (see, e.g., Eq.\ (53) in \cite{Sete2014}), which leads to correction of the effective drive amplitude,
    \be
    \tilde{\varepsilon} \approx \left[ 1-\frac{1}{2}\, (g/\Delta)^2\right] \varepsilon ,
    \label{eq:tilde-epsilon}\ee
within the ground-state eigenladder at $n\alt n_c$. (Within the excited-state eigenladder the correction will then be $\tilde{\varepsilon} \approx \{1+\frac{1}{2}(g/\Delta)^2 -\frac{1}{2} [\sqrt{2}\,g/(\Delta+\eta)]^2\}\, \varepsilon$.) Using the effective drive amplitude (\ref{eq:tilde-epsilon}) in Eqs.\ (\ref{eq:evol-W})--(\ref{eq:evol-beta}) instead of $\varepsilon$ produces dotted lines (instead of dashed lines) in Fig.\ \ref{fig:DSS}. We see that the dotted lines are quite close to the solid lines; therefore, the simple correction (\ref{eq:tilde-epsilon}) is sufficient for an accurate theory. Even better accuracy can be achieved if we use numerical matrix elements for the effective drive amplitude within the ground-state eigenladder,
    \be
    \tilde{\varepsilon} = \frac{\overline{\langle n-1,0|} a \overline{|n,0\rangle}}{\sqrt{n}} \, \varepsilon ,
    \ee
which now depends on $n\approx \bar{n}$. For Fig.\ \ref{fig:DSS} this produces a line (not shown), which closely follows the blue solid line for the squeezed-state approximation and a line practically indistinguishable from the orange solid line for the coherent-state approximation.

We emphasize that in Fig.\ \ref{fig:DSS} the infidelity of the dressed squeezed state model is $\alt 10^{-3}$, while for the dressed coherent state model it is only $\alt 10^{-1}$. Note that we always convert the sheared state (\ref{eq:sheared-Gaussian}) with parameters $K$ and $W$ into the squeezed state (\ref{eq:squeezed}) via Eqs.\ (\ref{eq:theta}) and (\ref{eq:r-conversion}) before comparing with numerical $|\psi\rangle_0$. If this is not done, the infidelity of the sheared Gaussian state in Fig.\ \ref{fig:DSS} would be above $10^{-3}$ at $t< 100$ ns ($\bar{n}<40$), reaching $3\times 10^{-2}$ for $\bar{n}<0.5$ and becoming practically equal to the blue lines only at $t>160$ ns ($\bar{n}>100$).

Thus, we have numerically confirmed that the dressed squeezed state approximation performs much better than the dressed coherent state approximation. Nevertheless, the inaccuracy of the dressed squeezed state model still grows in time, and may eventually become significant.

\section{Conclusion}\label{sec:conclusion}

In this paper we analyzed the ring-up of a readout resonator coupled to a transmon qubit. The bare bases of the transmon and resonator hybridize into a joint eigenbasis that is organized into natural eigenladders associated with each nominal transmon state. As was pointed out previously, ringing up the resonator from its ground state using a coherent pump approximately creates a coherent state in this eigenbasis (i.e., a dressed coherent state) that is confined to the eigenladder corresponding to the initial transmon state. We analyzed the deviations from this first approximation and developed a more accurate dynamical model for the ring-up process.

Through numerical simulation, we demonstrated that the ring-up evolution deviates from the dressed coherent state model in two important respects. First, the initial transmon population may leak into other (``incorrect'') eigenladders that correspond to different initial transmon states. Second, even within the initial (``correct'') eigenladder the state may differ from a coherent state. We analyzed both deviations and developed analytical models to quantify the effects.

The stray population that leaks outside the correct eigenladder arises from the mismatch between the coherent pump (in the bare basis) and the hybridized resonator (in the eigenbasis). We found that this mismatch creates interesting dynamics over a relatively short timescale after the pump is applied, and were able to describe the resulting damped oscillations between neighboring eigenladders quantitatively. The most important result is that for typical experimental parameters the occupation of incorrect eigenladders remains small ($\alt 10^{-4}$); therefore, this effect should not significantly contribute to the qubit measurement error in present-day experiments.  Note, however, that our analysis focuses solely on the population leakage caused by the pump itself during the ring-up process; as such, it neglects other important effects that contribute to the total leakage to incorrect eigenladders in practice, such as qubit energy relaxation, the Purcell effect, interactions with defects, dressed dephasing, and non-RWA effects. (Note that \cite{Sank2016} extends the analysis presented here to include non-RWA effects, thus explaining an important example of experimentally observed leakage at high photon numbers.)

The dynamics of the hybridized resonator state remaining within the correct eigenladder is non-trivial due to the effective resonator nonlinearity induced by the interaction with the transmon. This nonlinearity leads to a significant deviation from the dressed coherent state picture---in our numerical simulations the infidelity of the dressed coherent state reaches ${\sim}10^{-1}$. The nonlinear evolution shears the phase-space profile of the resonator state, deforming initially circular coherent state profiles into elliptical and crescent-shaped profiles over time. We showed that for practical ranges of parameters, these sheared profiles approximate ideal \emph{squeezed states in the eigenbasis} (i.e., dressed squeezed states)---in our simulations the infidelity of the squeezed state picture reaches ${\sim}10^{-3}$, or roughly two orders of magnitude better than that of a dressed coherent state. (Note that the dressed squeezed state is practically unentangled, similar to the dressed coherent state.) Using a hybrid phase-Fock-space approach, we derived simple equations of motion [Eqs.~\eqref{eq:evol-W}--\eqref{eq:evol-beta}] for the self-developing squeezing, which naturally generalize the evolution of a coherent state. These equations of motion depend only on the photon number-dependence of the dressed resonator frequency, which may be added phenomenologically from precomputed numerical simulations or measured experimentally.

We emphasize that the self-developing squeezing may significantly affect the qubit measurement error, either decreasing or increasing it, depending on the squeezing axis angle relative to the line passing through the state centers in the phase space for the qubit states $|0\rangle$ and $|1\rangle$. (The resonator field for the qubit state $|0\rangle$ is affected by squeezing much more than for the state $|1\rangle$ because of much more efficient level repulsion within the ground-state ladder of the Jaynes-Cummings Hamiltonian for the multi-level transmon.) Further analysis of this subject is definitely important.

The dressed squeezed state model provides an efficient and accurate description of the resonator physics during a sufficiently rapid ring-up process, when the resonator decay may be neglected (as was assumed in this paper). This regime is also physically relevant for at least two known protocols: the catch-disperse-release measurement of a qubit \cite{Sete2013} and the readout protocol \cite{Govia2014} based on Josephson photomultipliers. However, in the standard method of transmon measurement, the resonator decay cannot be neglected (except during the ring-up), that will require an extension of our dressed squeezed state model. This generalization will be considered in future work.

\begin{acknowledgments}
This research was supported by ARO grants W911NF-15-1-0496 and W911NF-11-1-0268. J.D. also acknowledges partial support by Perimeter Institute for Theoretical Physics. Research at Perimeter Institute is supported by the Government of Canada through Industry Canada and by the Province of Ontario through the Ministry of Economic Development \& Innovation.

\end{acknowledgments}

\appendix*

\section{Vanishing entanglement in dressed coherent and squeezed states} \label{app:unentangled}

In this appendix we show that dressed coherent states and dressed squeezed states are practically unentangled for large average numbers of photons $\bar{n}$. For a dressed coherent state we can anticipate this result because coherent states with large $\bar{n}$ are practically classical. Thus, the transmon is essentially driven by a classical field, and should therefore produce an unentangled state. However, this result is rather paradoxical because the dressed coherent state (\ref{eq:DCS}) is constructed out of highly entangled eigenstates of the transmon-resonator system, so significant entanglement could be naively expected. The derivation below resolves this paradox. A similar result also applies to a dressed squeezed state.

Let us consider a general dressed state
\begin{equation}\label{eq:dressed_state}
	\ket{\psi} = \sum\nolimits_n c_n \,\overline{\ket{n,k}},
\end{equation}
where  $\overline{\ket{n,k}}$ are the eigenstates of the transmon-resonator system for the transmon nominally in the state $|k\rangle_{\rm q}$, and the coefficients $c_n$ describe the nominal resonator state $\sum_n c_n \,\ket{n}_{\rm r}$, $\sum_n |c_n|^2=1$. Our first goal is to derive a condition for which this dressed state can be approximately represented as a direct product of the resonator state $\sum_n c_n \,\ket{n}_{\rm r}$ and some transmon state (which will be generally different from the nominal state $|k\rangle_{\rm q}$).

The eigenstate $\overline{\ket{n,k}}$ can be expanded in the bare basis (within the RWA strip) as
\begin{equation}\label{eq:eigen-expansion}
	\overline{\ket{n,k}} = \sum\nolimits_l d_l^{(n,k)} \,\ket{n-l,k+l},
\end{equation}
where the summation involves a few transmon levels, $-k\leq l\leq k_{\rm max}-k$, $k< k_{\rm max} \simeq 7$. The coefficients $d_l^{(n,k)}$ depend on $n$ because the coupling (\ref{eq:H_I}) between neighboring bare levels $|n-l,k+l\rangle$ and $|n-l-1,k+l+1\rangle$ is proportional to $\sqrt{n-l}$. However, this dependence can be neglected, $\sqrt{n-l}\approx \sqrt{\bar{n}-l}$ if
 	\begin{equation}
    \sigma_n \ll \bar{n}, \,\,\, k_{\rm max}\ll \bar{n},
    \label{eq:large-nbar}\end{equation}
where by the standard deviation $\sigma_n$ we characterize the spread of $n$ in the state (\ref{eq:dressed_state}).  In this case we can use approximation with $n$-independent coefficients $d_l^{(k)}$ (which may still depend on $\bar{n}$),
\begin{equation}\label{eq:eigen-approx}
	\overline{\ket{n,k}} \approx  \sum\nolimits_l d_l^{(k)} \,\ket{n-l,k+l}.
\end{equation}
Substituting Eq.\ (\ref{eq:eigen-approx}) into Eq.~\eqref{eq:dressed_state}, shifting the indices, $n-l \rightarrow n$, and changing the order of summation, we obtain
\begin{eqnarray}\label{eq:dressed_middle}
&&	\ket{\psi} \approx \sum\nolimits_l d_l^{(k)} \, \sum\nolimits_n c_{n+l} \,\ket{n,k+l}
	\nonumber \\
&& \hspace{1cm} = \sum\nolimits_l d_l^{(k)} \,\ket{k+l}_{\rm q} \, |\phi_l \rangle , \qquad
	\\
&&  |\phi_l \rangle = \sum\nolimits_n c_{n+l} \,\ket{n}_{\rm r},
\label{eq:phi-l-def}\end{eqnarray}
where $|k+l\rangle_{\rm q}$ is the transmon level and $|\phi_l \rangle$ is the resonator state, which depends on transmon index $l$. Note that $|\phi_l \rangle$  are (practically) normalized, since the coefficients $c_{n+l}$ are the same as in the normalized state (\ref{eq:dressed_state}) and the shift of indices by $l$ is not important when the condition (\ref{eq:large-nbar}) is satisfied.

The dependence of $|\phi_l \rangle$ on the transmon index $l$ indicates the entanglement between the transmon and resonator. If $|\phi_l \rangle$ were not dependent on $l$, then $|\psi\rangle$ in Eq.\ (\ref{eq:dressed_middle}) is an (unentangled) direct product of the transmon and resonator states. Moreover, any $l$-dependent phase factor, $|\phi_l \rangle = e^{i\varphi_l} |\phi_0\rangle$, may be absorbed into the transmon state, still yielding a direct product. This gives us a {\it condition} for the approximate absence of entanglement: $|\langle \phi_0 |\phi_l\rangle |\approx 1$ for all transmon indices $l$.

Thus, we have shown that if
	\begin{equation}
   \bigg| \sum\nolimits_n c_n^* c_{n+l} \bigg| \approx 1
    \label{eq:condition}\end{equation}
for any $l$ within the relevant range, $|l|\leq k_{\rm max}\simeq 7$, then the dressed state (\ref{eq:dressed_state}) is approximately a direct product,
	\begin{equation}
  \sum\nolimits_n c_n \,\overline{\ket{n,k}} \approx \sum\nolimits_n c_n \,\ket{n}_{\rm r} \otimes \sum\nolimits_l  e^{i\varphi_l} d_l^{(k)} |k+l\rangle_{\rm q} ,
    \label{eq:direct-product}\end{equation}
where $\varphi_l={\rm arg}(\sum_n  c_n^* c_{n+l})$ and $d_l^{(k)}$ are the coefficients in the eigenstate (\ref{eq:eigen-approx}).

Now let us show that the condition (\ref{eq:condition}) is satisfied for a {\it  dressed coherent state} $|\alpha\rangle_k$ given by Eq.\ (\ref{eq:DCS}). Since in this case $c_n = \exp(-|\alpha|^2/2)\, \alpha^n/\sqrt{n!}$, we find
\begin{align}
	\sum_n c_n^* c_{n+l}  &= \sum_n e^{-|\alpha|^2} \frac{|\alpha|^{2n}}{n!} \frac{|\alpha|^l e^{il \, {\rm arg} (\alpha)}}{\sqrt{(n+1)(n+2)\dotsm(n+l)}}
    \nonumber \\
    &\approx e^{i\varphi_l}, \,\,\, \varphi_l =l\,\mathrm{arg}(\alpha),
\end{align}
where we approximated $\sqrt{(n+1)(n+2)\dotsm(n+l)}\approx n^{l/2}\approx |\alpha|^l$. This approximation requires $|\alpha |^2\gg l^2$. Thus, the dressed coherent state $|\alpha\rangle_k$ is practically unentangled if $|\alpha|^2 \gg k_{\rm max}^2$.

\begin{figure}[tb]
	\begin{center}
		\includegraphics[width=7.5cm]{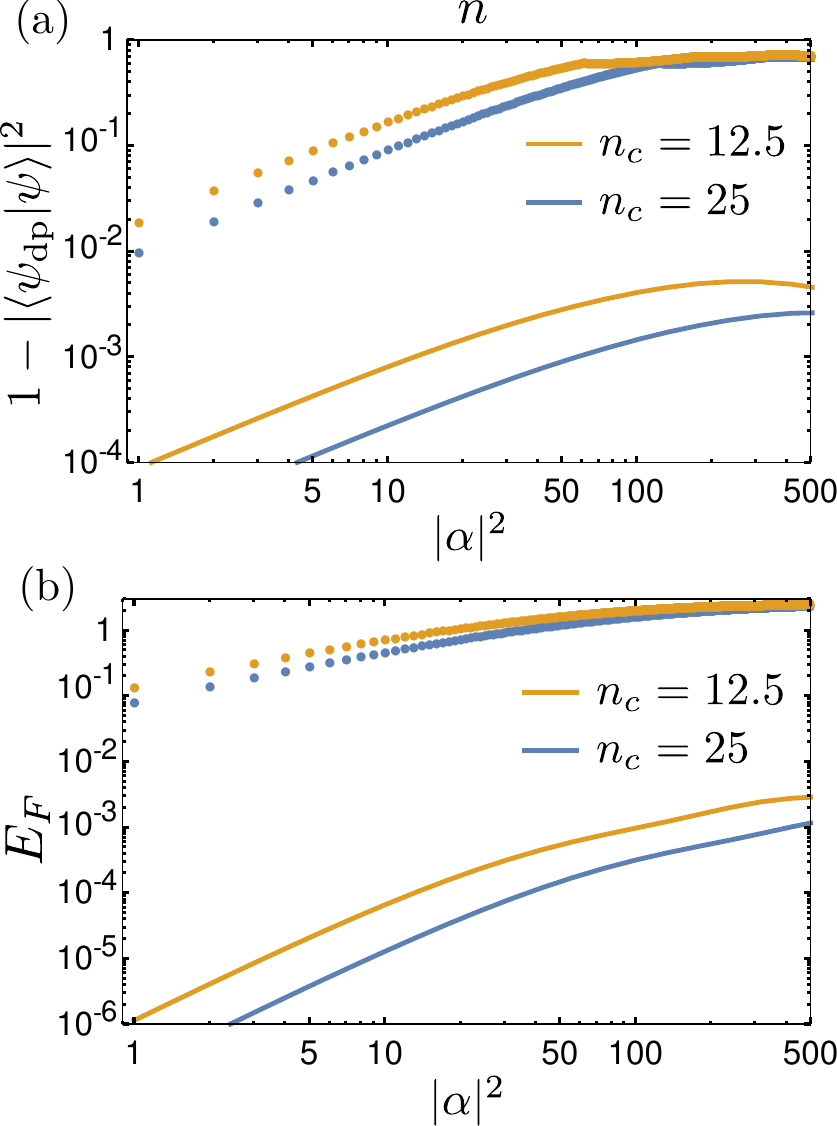}
    \end{center}
    \caption{(a) Solid lines: infidelity $1-|\langle\psi_{\rm dp}|\alpha\rangle_0|^2$ of approximating the dressed coherent state $|\alpha\rangle_0$ with a direct-product state $|\psi_{\rm dp}\rangle$ given by Eq.\ (\ref{eq:direct-product}), as a function of $|\alpha|^2$. For comparison, the dots show similar infidelity for the eigenstates  $\overline{|n,0\rangle}$, i.e., dressed Fock states, as a function of $n$ (axes of $n$ and $|\alpha|^2$ coincide). We assume $(\omega_{\rm r}-\omega_{\rm q})/2\pi = 1$ GHz,  $\eta/2\pi=200$ MHz, and $g/2\pi=100$ MHz (lower blue line/dots, $n_c=25$) or $g/2\pi =141.4$ MHz (upper orange line/dots, $n_c=12.5$). (b) Entanglement of formation $E_F$ (coinciding with entropy of entanglement) for the dressed coherent states $|\alpha\rangle_0$ (lines) and dressed Fock states (dots) with the same parameters as in (a).
    } \label{fig:direct-product}
\end{figure}

The solid lines in Fig.\ \ref{fig:direct-product}(a) show the inaccuracy of the direct-product approximation (\ref{eq:direct-product}) for the dressed coherent state $|\alpha\rangle_0$ as a function of $|\alpha|^2$ for typical parameters: $(\omega_{\rm r}-\omega_{\rm q})/2\pi = 1$ GHz,  $\eta/2\pi=200$ MHz, and $g/2\pi=100$ MHz (lower blue line, $n_c=25$) or $g/2\pi =141.4$ MHz (upper orange line, $n_c=12.5$). As a measure of inaccuracy we use  $1-|\langle\psi_{\rm dp}|\alpha\rangle_0|^2$, where the direct-product state $|\psi_{\rm dp}\rangle$ is given by Eq.\ (\ref{eq:direct-product}). Note that for small $\alpha$ we average coefficients $d_l^{(n,k)}$ in Eq.\ (\ref{eq:eigen-expansion}) to obtain $d_l^{(k)}$. We see that the solid lines in Fig. \ref{fig:direct-product}(a) significantly increase with $\bar{n}\approx|\alpha|^2$ until $\bar{n}$ becomes much larger than $n_c$. This behavior is due to a competition between the continuously increasing entanglement of eigenstates (\ref{eq:eigen-expansion}) and the decrease of entanglement due to the increasingly satisfied condition (\ref{eq:condition}). However, we see that even at large $\bar{n}$, the dressed coherent state $|\alpha\rangle_0$ is very close to the direct-product state (\ref{eq:direct-product}). For comparison, we show with blue and orange dots the much larger inaccuracy when we try to approximate the corresponding eigenstates $\overline{|n,0\rangle}$ (i.e., the dressed Fock states) with similar direct-product wavefunctions. It is easy to prove that the best such approximation is the bare state with the largest coefficient in the expansion (\ref{eq:eigen-expansion}); the visible kinks in Fig.\ \ref{fig:direct-product}(a) are due to the change of this best bare state. Figure \ref{fig:direct-product}(b) is similar to Fig.\ \ref{fig:direct-product}(a), except it shows the entanglement of formation \cite{Plenio2014} (equal to the entropy of entanglement for pure states) for the same dressed coherent states $|\alpha\rangle_0$ and dressed Fock states $\overline{|n,0\rangle}$. With this measure we again confirm  that the dressed coherent states are practically unentangled, in contrast to the strongly entangled dressed Fock states [note an overall similarity between Figs.\ \ref{fig:direct-product}(a) and \ref{fig:direct-product}(b)].

Even though a dressed coherent state is practically unentangled, there is a strong classical correlation between the resonator and transmon dynamics. This can be seen by adding explicit time dependence into Eq.\ (\ref{eq:direct-product}), thus going from the rotating frame into the lab frame. Replacing coefficients $c_n$ for the coherent state with $c_n(t)= e^{-in\omega_{\rm r}t}c_n(0)$ (the remaining factor $e^{-iE(k,\bar{n})t}$ is an overall phase and therefore not important), we find $\alpha (t)= e^{-i\omega_{\rm r}t} \alpha (0)$. As a result, $\phi_l=l \, {\rm arg} [\alpha(0)] -l\omega_{\rm r}t$, and therefore the dressed coherent state evolves in time as
	\begin{equation}
|\alpha \rangle_k = |e^{-i\omega_{\rm r}t}\alpha(0)\rangle_{\rm r} \otimes \sum_l e^{-il\omega_{\rm r}t} e^{il{\rm arg}[\alpha(0)]}  d_l^{(k)}    |k+l\rangle_{\rm q} .
    \label{eq:qubit-state-t}\end{equation}
We see that both resonator and transmon states are evolving with the period $2\pi/\omega_{\rm r}$ in a phase-synchronized way; the resonator state evolution is a simple oscillation, but the transmon evolution within the period is quite non-trivial. This is exactly what we would expect classically for a non-linear oscillator that is harmonically driven with frequency $\omega_{\rm r}$. We have performed numerical simulations for the transmon state evolution in Eq.\ (\ref{eq:qubit-state-t}) using the $x$-representation (where $x$ in this case is the superconducting phase difference) and confirmed such non-trivial evolution within one period of oscillations when $\bar{n}$ is significantly larger than $n_c$.

To check the direct-product condition (\ref{eq:condition}) for a {\it dressed squeezed state}, let us use its approximate sheared Gaussian representation in Eq.~\eqref{eq:sheared-Gaussian}. Then we find
	\begin{equation}
    \sum_n c_n^*c_{n+l} \approx e^{il \, {\rm arg(\beta)}}\left[ 1-\frac{l^2}{2W\bar{n}} -\frac{2K^2W}{\bar{n}}\,l^2\right] ,
    \end{equation}
assuming large $\bar{n}= |\beta|^2$. We see that the condition (\ref{eq:condition}) is satisfied if
	\begin{equation}
    \bar{n}\gg k_{\rm max}^2 \max (1/2W,\, 2K^2W).
    \end{equation}
In this case the dressed squeezed state is practically unentangled. For the dressed coherent state ($W=1$, $K=0$) this inequality reduces to  $\bar{n}\gg k_{\rm max}^2$, as expected.

Note that in the case when the dressed sheared state is practically unentangled, the phase $\varphi_l=l\, {\rm arg (\beta)}$ in Eq.\ (\ref{eq:direct-product}) is still the same as for the dressed coherent state (except for the notation change, $\alpha \rightarrow \beta$). Therefore, the transmon state and its evolution within the period of $\omega_{\rm r}$ is still the same as for the dressed coherent state with $\alpha=\beta$. In other words, for sufficiently large $\bar{n}$ there is no difference for the transmon if it is driven by a coherent or a squeezed field from the resonator.


%

\end{document}